\documentclass[a4paper,english,titlepage,12pt]{article}

\usepackage{babel}\usepackage{amsmath,amssymb,amsthm}
\usepackage[numbers,compress]{natbib}
\usepackage{graphicx}
\usepackage{longtable}
\usepackage{float}
\usepackage{titlesec}
\usepackage{caption}

\makeatletter
\def\@xfootnote[#1]{%
  \protected@xdef\@thefnmark{#1}%
  \@footnotemark\@footnotetext}
\makeatother

\newcommand{\new}{\text{\textit{\textbf{new}}}}
\newcommand{\rop}{\mathcal{R}}
\newcommand{\ffdeg}{\mathrm{Deg}}
\newcommand{\forests}{\mathfrak{F}}
\newcommand{\infragr}{\mathfrak{I}}
\newcommand{\forestsmax}{\mathfrak{F}_{\text{max}}}
\newcommand{\iclos}{\mathrm{IClos}}
\newcommand{\electrons}{e}
\newcommand{\sechains}{\mathrm{SE}}
\newcommand{\cbigz}{C_{\text{bigZ}}}
\newcommand{\cbigf}{C_{\text{bigF}}}
\newcommand{\csubse}{C_{\text{subSE}}}
\newcommand{\csubi}{C_{\text{subI}}}
\newcommand{\csubo}{C_{\text{subO}}}
\newcommand{\cadd}{C_{\text{add}}}
\newcommand{\ffsub}{\mathrm{Sub}}
\newcommand{\rndu}{^{\text{up}}}
\newcommand{\rndd}{^{\text{down}}}
\newcommand{\appr}{^{\text{appr}}}
\newcommand{\rnd}{^{\text{rnd}}}
\newcommand{\maxlog}{\mathrm{maxlog}}
\newcommand{\uncert}{_{\text{uncert}}}
\newcommand{\peak}{_{\text{peak}}}
\newcommand{\argmin}{\mathrm{argmin}}

\renewcommand{\thesection}{\Roman{section}}

\makeatletter
\renewcommand{\p@subsection}{\thesection.}
\makeatother
\titlelabel{\thetitle.\ }

\begin{document}

\begin{center}\LARGE{\textbf{Numerical calculation of high-order QED contributions to the electron
anomalous magnetic moment}}
\end{center}
\begin{center}
\large{Sergey Volkov\footnote[*]{E-mail: \texttt{volkoff\underline{
}sergey@mail.ru, sergey.volkov.1811@gmail.com}}}
\\ \ \\ \emph{\small{SINP MSU, Moscow, Russia \\ DLNP JINR, Dubna, Russia}}
\end{center}

\small{This paper describes a method of numerical evaluating
high-order QED contributions to the electron anomalous magnetic
moment. The method is based on subtraction of infrared and
ultraviolet divergences in Feynman-parametric space before
integration and on nonadaptive Monte Carlo integration that is
founded on Hepp sectors. A realization of the method on the graphics
accelerator NVidia Tesla K80 is described. A method of removing
round-off errors that emerge due to numerical subtraction of
divergences without losing calculation speed is presented. The
results of applying the method to all 2-loop, 3-loop, 4-loop QED
Feynman graphs without lepton loops are presented. A detailed
comparison of the 2-loop and 3-loop results with known analytical
ones is given in the paper. A comparison of the contributions of 6
gauge invariant 4-loop graph classes with known analytical values is
presented. Moreover, the contributions of 78 sets of 4-loop graphs
for comparison with the direct subtraction on the mass shell are
presented. Also, the contributions of the 5-loop and 6-loop ladder
graphs are given as well as a comparison of these results with known
analytical ones. The behavior of the generated Monte Carlo samples
is described in detail, a method of the error estimation is
presented. A detailed information about the graphics processor
performance on these computations and about the Monte Carlo
convergence is given in the paper. } \normalsize

\section{INTRODUCTION}\label{sec_intro}

The electron anomalous magnetic moment (AMM) is known with a very
high accuracy. In the experiment~\cite{experiment} the value
$$
a_e=0.00115965218073(28)
$$
was obtained. So, an extremely high precision is required also from
theoretical predictions.

The most precise prediction of electron's AMM at the present time
 uses the following representation:
$$
a_e=a_e(\text{QED})+a_e(\text{hadronic})+a_e(\text{electroweak}),
$$
$$
a_e(\text{QED})=\sum_{n\geq 1} \left(\frac{\alpha}{\pi}\right)^n
a_e^{2n},
$$
$$
a_e^{2n}=A_1^{(2n)}+A_2^{(2n)}(m_e/m_{\mu})+A_2^{(2n)}(m_e/m_{\tau})+A_3^{(2n)}(m_e/m_{\mu},m_e/m_{\tau}),
$$
where $m_e$, $m_{\mu}$, $m_{\tau}$ are masses of electron, muon, and
tau lepton, respectively. Different terms of this expression were
calculated by different groups of researchers. Some of them has
independent calculations, but some of them were calculated only by
one scientific group. The best theoretical value
~\cite{kinoshita_10_corr}
\begin{equation}\label{eq_kinoshita_amm}
a_e=0.001159652182032(13)(12)(720)
\end{equation}
was obtained by using the fine structure constant
$\alpha^{-1}=137.035998995(85)$ that had been obtained by using
independent from $a_e$ methods (see ~\cite{kinoshita_10_corr}).
Here, the first, second, and third uncertainties come from
$A_1^{(10)}$, $a_e(\text{hadronic})+a_e(\text{electroweak})$ and the
fine-structure constant\footnote{So, the calculated coefficients are
used for improving the accuracy of $\alpha$.} respectively. The
values
$$
A^{(2)}_1=0.5,
$$
$$
A_1^{(4)}=-0.328478965579193\ldots,
$$
$$
A_1^{(6)}=1.181241456\ldots,
$$
$$
A_1^{(8)}=-1.9122457649\ldots
$$
are known from the analytical and semianalytical results in
~\cite{schwinger1,schwinger2}, ~\cite{analyt2_p,analyt2_z},
~\cite{analyt3}, ~\cite{laporta_8}, respectively\footnote{The value
for $A_1^{(6)}$ was a product of efforts of many scientists. See,
for example,
~\cite{analyt_mi,analyt_b2,analyt_b3,analyt_b1,analyt_b4,
analyt_b,analyt_e,analyt_d,analyt_c,analyt_ll,analyt_f,analyt_g,analyt_h,
analyt_i, analyt_j, laporta_1993}.}. The value
$$
A_1^{(10)}=6.675(192)
$$
was presented in ~\cite{kinoshita_10_corr}.
 At the present time, there are no independent calculations
of $A_1^{(10)}$. However, $A_1^{(8)}$ was evaluated
independently\footnote{However, by 2016 most part of $A_1^{(8)}$ had
been calculated by only one scientific group
~\cite{kinoshita_8_last}. First numerical estimations for
$A_1^{(8)}$ were presented in ~\cite{kinoshita_8_first}.} in
~\cite{kinoshita_8_last,rappl,smirnov_amm} (and for the graphs
without lepton loops in ~\cite{volkov_prd}). We must take into
account the fact that the contributions of some individual graphs
turn out to be several times greater than the total contribution in
absolute value\footnote{It turns out regardless of the used
divergence subtraction method.}. Therefore, an error in one graph
evaluation can make the final result to be entirely wrong. So, the
problem of evaluating $A_1^{(2n)}$ is still relevant.

The most uncertain and difficult for evaluation QED contributions to
$a_e$ correspond to Feynman graphs without lepton loops. We consider
an evaluation of these contributions in this paper and denote the
$n$-loop part of it by $A_1^{(2n)}[\text{no lepton loops}]$.

This paper is a continuation of the series of papers
~\cite{volkov_2015,volkov_prd} with increasing precision, number of
independent loops in graphs, refinement of the consideration.

We use the subtraction procedure for removing both infrared and
ultraviolet divergences that was introduced in ~\cite{volkov_2015}.
It is briefly described in Section \ref{sec_subtraction}. This
procedure eliminates IR and UV divergences in each AMM Feynman graph
point-by-point, before integration, in the spirit of papers
~\cite{kinoshita_10_corr, levinewright, carrollyao, carroll,
kinoshita_6, kompaniets, bogolubovparasuk, hepp, zavialovstepanov,
scherbina, zavialov, smirnov} etc. This property is substantial for
many-loop calculations when reducing the computer time is of
critical importance. Let us note that $A_1^{(2n)}$ is free from
infrared divergences since they are removed by the on-shell
renormalization as well as the ultraviolet ones (see a more detailed
explanation in ~\cite{volkov_2015}). However, the subtractive
on-shell renormalization can't eliminate IR divergences in
Feynman-parametric space before integration as well as it does for
UV divergences\footnote{Moreover, it can generate additional
IR-divergences, see a more detailed explanation in
~\cite{volkov_2015}.}. The structure of IR and UV divergences in
individual Feynman graphs is quite complicated\footnote{See notes in
~\cite{volkov_prd}.}. Therefore, a special procedure is required for
removing both UV and IR divergences. Let us recapitulate the
advantages of the developed subtraction procedure.
\begin{enumerate}
\item It is fully automated for any $n$.
\item It is comparatively easy for realization on computers.
\item It can be represented as a forestlike formula. This formula differs from
the classical forest formula ~\cite{zavialovstepanov, scherbina,
zimmerman} only in the choice of linear operators and in the way of
combining them.
\item The contribution of each Feynman graph to $A_1^{(2n)}$ can be represented as a single
Feynman-parametric integral. The value of $A_1^{(2n)}$ is the sum of
these contributions.
\item Feynman parameters can be used directly, without any
additional tricks.
\end{enumerate}
See a detailed description in ~\cite{volkov_2015}. The subtraction
procedure was checked independently by F. Rappl using Monte Carlo
integration based on Markov chains ~\cite{rappl}. An additional
advantage of the procedure is described below and in Section
\ref{subsec_classes}.

After the subtraction is applied, the problem is reduced to
numerical integration of functions of many variables. The number of
variables can be quite big\footnote{For example, for $5$ loops we
have $13$ variables, see ~\cite{volkov_prd} and Section
\ref{subsec_real_overview}.}, this fact compels us to use Monte
Carlo methods. In most cases the precision of Monte Carlo
integration behaves asymptotically as $C/\sqrt{N}$, where $N$ is the
number of samples. Thus, for reaching a high precision in practical
time it is very important to decrease the constant $C$ as much as
possible. Unfortunately, the behavior of Feynman-parametric
integrands that appear in $A_1^{(2n)}$ computation often leads to
slow Monte Carlo convergence. An integration method with a
relatively good constant $C$ was introduced in ~\cite{volkov_prd}.
The method is based on importance sampling with probability density
functions that are constructed for each Feynman graph individually.
The construction is based on Hepp sectors ~\cite{hepp} and uses
functions of the form that was first used by E. Speer ~\cite{speer}
with some modifications. The modification is based on the concept of
I-closure that was introduced in ~\cite{volkov_prd}. The method from
~\cite{volkov_prd} demonstrated better convergence than the
universal Monte Carlo routines. A refined version of the
construction is described in Section \ref{sec_monte_carlo}. This
refinement reduces the uncertainty of $A_1^{(8)}[\text{no lepton
loops}]$ by about 15\% when the number of samples is fixed.

When we have a deal with unbounded functions or with functions
having sharp peaks, the standard Monte Carlo error estimation
approach has a tendency to underestimate the inaccuracy. A method of
preventing underestimation was described in ~\cite{volkov_prd}.
However, some tests show that in many cases a more accurate
consideration of peaks is required. An improved method of error
estimation that uses a specificity of the considering integrands is
presented in Section \ref{subsec_error_est}. A detailed information
about samples behavior for the 5-loop and 6-loop ladder graphs is
provided. Also, an information about the dependence of the results
on number of samples is given for $A_1^{(8)}[\text{no lepton
loops}]$ and for the 5-loop and 6-loop ladder graph contributions.

Numerical subtraction of divergences leads to the situation when
small numbers (in absolute value) are obtained as a difference of
astronomically big numbers. This generates round-off errors that
significantly affect the result\footnote{Moreover, these errors can
convert a finite result to an infinite one.}. To control these
errors we need to use additional techniques that substantially slows
down the computation speed. In ~\cite{volkov_2015} all integrand
evaluations were first performed with two different
precisions\footnote{in 64-bit in 80-bit precisions that are
supported on processors that are compatible with Intel x86 family},
and when the difference of the results was noticeable, the
calculation was repeated with increased precision. This approach
requires twice as much computer time than the direct calculation.
Also, an emergence of bias is possible in this case. All
calculations that are described in ~\cite{volkov_prd} use the
interval arithmetic\footnote{See Section \ref{subsec_ia}.}. The
interval arithmetic is reliable but slows down the computation many
times: for example, a multiplication of two intervals requires 8
number multiplications with correct rounding, 3 minimums and 3
maximums. To eliminate this slowdown a special modification of the
interval arithmetic was developed. This technique gave a significant
improvement in computation speed without loss of reliability. In
many cases this method works faster than the approach with two
precisions\footnote{See Table \ref{table_tech}.}. A specificity of
the construction of the integrands is used for reaching such a
performance. The description of this technique is contained in
Section \ref{subsec_eia}.

Rapid development of specialized computing devices that solve some
tasks many times faster than ordinary computers makes it possible to
use them for scientific calculations. All Monte Carlo integrations
that are described in this paper were performed on
one\footnote{NVidia Tesla K80 has 2 GPUs.} graphics processor of
NVidia Tesla K80. Graphics processors (GPUs) are very useful for
Monte Carlo integration. However, a specific programming is required
to use these devices effectively. Sections
\ref{subsec_real_overview}, \ref{subsec_prec}, \ref{subsec_tech}
contain some information about the realization of the described
integration method on GPU.

The developed method and realization were applied for computing
$A_1^{(2n)}[\text{no lepton loops}]$, $n=2,3,4$. Also, the
contributions of the 5-loop and 6-loop ladders were evaluated for
testing purposes. The results are presented in Section
\ref{subsec_real_overview}. The comparison with known analytical
results is provided in Table \ref{table_tech}.

High-order calculations in Quantum Field Theory require performing
some operations with enormous amounts of information. For example,
the total integrand code size\footnote{See Table \ref{table_tech}.}
for $A_1^{(8)}[\text{no lepton loops}]$ is 2.5 GB. There are too
many places where a mistake can emerge. However, the total
independent check requires a lot of resources. So, it is very
important to have a possibility to check the results by parts by
using another methods. Section \ref{subsec_classes} demonstrates
that the developed method provides such a possibility. The total set
of 269 Feynman graphs for $A_1^{(8)}[\text{no lepton loops}]$ is
divided into 78 subsets, the contribution of each set must coincide
with the contribution that is obtained by direct subtraction on the
mass shell in Feynman gauge. The contribution of each set is
provided in Section \ref{subsec_classes}. Also, an analogous
information is given for the 2-loop and 3-loop cases, the comparison
with known analytical results is provided as well as it has been
done in ~\cite{volkov_2015}. The contributions of 6 gauge invariant
classes of 4-loop graphs without lepton loops are presented in
Section \ref{subsec_classes} and compared with the semianalytical
ones from ~\cite{laporta_8}. Knowing the values of contributions of
gauge invariant classes gives an ability to check some hypotheses
from Quantum Field Theory\footnote{See Section
\ref{sec_conclusion}.}. Section \ref{subsec_indi} contains the
detailed information about contributions of individual Feynman
graphs including the influence of round-off errors and the
information about Monte Carlo error estimation. The summary of the
results and the technical information about GPU performance and code
sizes is presented in Section \ref{subsec_tech}.

\section{SUBTRACTION OF DIVERGENCES}\label{sec_subtraction}

We will work in the system of units, in which $\hbar=c=1$, the
factors of $4\pi$ appear in the fine-structure constant:
$\alpha=e^2/(4\pi)$, the tensor $g_{\mu\nu}$ is defined by
$$
g_{\mu\nu}=g^{\mu\nu}=\left(\begin{matrix}1 & 0 & 0 & 0 \\ 0 & -1 &
0 & 0 \\ 0 & 0 & -1 & 0 \\ 0 & 0 & 0 & -1 \end{matrix}\right),
$$
the Dirac gamma-matrices satisfy the condition
$\gamma^{\mu}\gamma^{\nu}+\gamma^{\nu}\gamma^{\mu}=2g^{\mu\nu}$.

We will use Feynman graphs with the propagators
\begin{equation}\label{eq_electron_propagator}
\frac{i(\hat{p}+m)}{p^2-m^2+i\varepsilon}
\end{equation} for
electron lines and
\begin{equation}\label{eq_feynman_gauge}
\frac{-g_{\mu\nu}}{p^2+i\varepsilon}
\end{equation}
for photon lines. We restrict our attention to graphs without lepton
loops. However, the developed subtraction procedure works for graphs
with lepton loops as well ~\cite{volkov_2015}.

The number $\omega(G)=4-N_{\gamma}-\frac{3}{2}N_e$ is called the
\emph{ultraviolet degree of divergence} of the graph $G$. Here,
$N_{\gamma}$ is the number of external photon lines of $G$, $N_e$ is
the number of external electron lines of $G$.

A subgraph\footnote{In this paper we take into account only such
subgraphs that are strongly connected and contain all lines that
join the vertexes of the given subgraph.} $G'$ of the graph $G$ is
called UV-divergent if $\omega(G')\geq 0$. There are the following
types of UV-divergent subgraphs in QED Feynman graphs without lepton
loops: \emph{electron self-energy subgraphs} ($N_e=2,N_{\gamma}=0$)
and \emph{vertexlike} subgraphs ($N_e=2,N_{\gamma}=1$).

Two subgraphs are said to overlap if they are not contained one
inside the other, and their sets of lines have a non-empty
intersection.

A set of subgraphs of a graph is called a \emph{forest} if any two
elements of this set don't overlap.

For a vertexlike graph $G$ by $\forests[G]$ we denote the set of all
forests $F$ consisting of UV-divergent subgraphs of $G$ and
satisfying the condition $G\in F$. By $\infragr[G]$ we denote the
set of all vertexlike subgraphs $G'$ of $G$ such that $G'$ contains
the vertex that is incident\footnote{We say that a line $l$ and a
vertex $v$ are \emph{incident} if $v$ is one of the endpoints of
$l$.} to the external photon line of $G$.\footnote{In particular,
$G\in \infragr[G]$.}

We will use the following linear operators that are applied to the
Feynman amplitudes of UV-divergent subgraphs:
\begin{enumerate}
\item $A$ is the projector of AMM. This
operator is applied to the Feynman amplitudes of vertexlike
subgraphs. See the definition in ~\cite{volkov_2015,volkov_prd}.
\item The definition of the operator $U$ depends on the type of
UV-divergent subgraph to which the operator is applied:
\begin{itemize}
\item
If $\Sigma(p)$ is the Feynman amplitude that corresponds to an
electron self-energy subgraph,
\begin{equation}\label{eq_sigma_general}
\Sigma(p)=u(p^2)+v(p^2)\hat{p},
\end{equation}
then, by definition\footnote{Note that it differs from the standard
on-shell renormalization.},
\begin{equation}\label{eq_u_self_energy}
U\Sigma(p) = u(m^2)+v(m^2)\hat{p}.
\end{equation}
\item If $\Gamma_{\mu}(p,q)$ is the Feynman amplitude\footnote{These rules are applied for individual
Feynman graphs and even for fixed values of Feynman parameters. So,
we can not neglect $\ldots(\hat{p}\gamma_{\mu}-\gamma_{\mu}\hat{p})$
terms, we can not use the Ward-Takahashi identity or other
simplifications.} corresponding to a vertexlike subgraph,
\begin{equation}\label{eq_gamma_general_q0}
\Gamma_{\mu}(p,0)=a(p^2)\gamma_{\mu} + b(p^2)p_{\mu} +
c(p^2)\hat{p}p_{\mu}+d(p^2)(\hat{p}\gamma_{\mu}-\gamma_{\mu}\hat{p}),
\end{equation}
then, by definition,
\begin{equation}\label{eq_u_vertex}
U\Gamma_{\mu}=a(m^2) \gamma_{\mu}.
\end{equation}
\end{itemize}
The operator $U$ can be used for extracting the UV-divergent part of
the amplitude without touching the IR-divergent part. For example,
for the one-loop amplitude (\ref{eq_gamma_general_q0}) all UV
divergences are contained in $a(p^2)\gamma_{\mu}$, but all IR
divergences are in $b(p^2)p_{\mu} + c(p^2)\hat{p}p_{\mu}$. For the
one-loop amplitude (\ref{eq_sigma_general}) IR divergences appear
after on-shell differentiating that is needed in the standard
renormalization, but not for defining $U$. See a detailed
description in terms of Feynman parameters in ~\cite{volkov_2015}.
It is important that $U$ preserves the Ward identity. This fact is
used for proving that the subtraction procedure is equivalent to the
on-shell renormalization and for calculating the contributions of
graph classes, see ~\cite{volkov_2015} and Section
\ref{subsec_classes}. It is also important for removing IR
divergences that (\ref{eq_u_self_energy}) extracts the self-mass
completely, see Discussion in ~\cite{volkov_2015}.
\item $L$ is the operator that is used in the standard subtractive on-shell renormalization
of vertexlike subgraphs. If $\Gamma_{\mu}(p,q)$ is the Feynman
amplitude that corresponds to a vertexlike subgraph,
(\ref{eq_gamma_general_q0}) is satisfied, then, by definition,
\begin{equation}\label{eq_q_def}
L\Gamma_{\mu}=[a(m^2)+mb(m^2)+m^2c(m^2)]\gamma_{\mu}.
\end{equation}
\end{enumerate}

Let $f_G$ be the unrenormalized Feynman amplitude that corresponds
to a vertexlike graph $G$. Let us write the symbolic definition
\begin{equation}\label{eq_rop_tilde}
\tilde{f}_G=\rop^{\new}_G f_G,
\end{equation}
where
\begin{equation}\label{eq_rop}
\rop^{\new}_G=\sum_{\substack{F=\{G_1,\ldots,G_n\}\in \forests[G] \\
G'\in \infragr[G]\cap F}}(-1)^{n-1}M^{G'}_{G_1}M^{G'}_{G_2}\ldots
M^{G'}_{G_n},
\end{equation}
\begin{equation}\label{eq_operators}
M^{G'}_{G''}=\begin{cases}A_{G'},\text{ if }G'=G'', \\
U_{G''},\text{ if }G''\notin \infragr[G]\text{, or }G''\varsubsetneq
G',
\\ L_{G''},\text{ if }G''\in \infragr[G], G'\varsubsetneq G'', G''\neq
G,
\\ (L_{G''}-U_{G''}),\text{ if }G''=G, G'\neq G.\end{cases}
\end{equation}
In this notation, the subscript of an operator symbol denotes the
subgraph to which this operator is applied.

The coefficient before $\gamma_{\mu}$ in $\tilde{f}_G$ is the
contribution of $G$ to $a_e$. See the examples of applying the
procedure in ~\cite{volkov_2015,volkov_prd}. The operators $L_{G''}$
and $(L_{G''}-U_{G''})$ are used for removing the IR divergences
that are connected with subgraphs in the sense of
~\cite{kinoshita_infrared} and the corresponding UV ones. Note that
the operator $(L_{G''}-U_{G''})$ is required in (\ref{eq_operators})
for removing UV divergences\footnote{See ~\cite{volkov_2015},
Appendix C.} and to make this subtraction to be equivalent to the
on-shell renormalization\footnote{See Section ~\ref{subsec_classes}
and ~\cite{volkov_2015} (Appendix B).}, it can not be replaced by
$L_{G''}$.

\section{PROBABILITY DENSITY FUNCTIONS FOR MONTE CARLO INTEGRATION}\label{sec_monte_carlo}

We use Feynman parameters for calculations. Thus, to obtain the
contribution of a graph $G$ we need to calculate the integral
$$
\int_{z_1,\ldots,z_n>0}I(z_1,\ldots,z_n)\delta(z_1+\ldots+z_n-1)dz_1\ldots
dz_n,
$$
where the function $I$ is constructed by using the known rules
~\cite{volkov_2015}.

We use the Monte Carlo approach based on importance sampling: we
generate randomly $N$ samples
$\underline{z}_1,\ldots,\underline{z}_N$, where
$\underline{z}_j=(z_{j,1},\ldots,z_{j,n})$, using some probability
density function $g(\underline{z})$ and approximate the integral
value by
\begin{equation}\label{eq_integral_appr}
\frac{1}{N}\sum_{j=1}^N \frac{I(\underline{z})}{g(\underline{z})}.
\end{equation}

The density $g$ is fixed for a fixed graph $G$. The speed of Monte
Carlo convergence depends on selection of $g$. A construction of $G$
that gives a good convergence is described below.

We will use Hepp sectors ~\cite{hepp} and functions of the form that
was first used by E. Speer ~\cite{speer} with some modifications.
All the space $\mathbb{R}^n$ is split\footnote{Let us remark that
the components has intersections on their boundaries. However, this
is inessential for integration.} into \emph{sectors}. Each sector
corresponds to a permutation $(j_1,\ldots,j_n)$ of
$\{1,2,\ldots,n\}$ and is defined by
$$
S_{j_1,\ldots,j_n}=\{(z_1,\ldots,z_n)\in\mathbb{R}:\ z_{j_1}\geq
z_{j_2}\geq\ldots\geq z_{j_n}\}.
$$
We define the function $g_0(z_1,\ldots,z_n)$ on $S_{j_1,\ldots,j_n}$
by the following relation
\begin{equation}\label{eq_density_0}
g_0(z_1,\ldots,z_n)=\frac{\prod_{l=2}^n
(z_{j_l}/z_{j_{l-1}})^{\ffdeg(\{j_l,j_{l+1},\ldots,j_n\})}}{z_1z_2\ldots
z_n},
\end{equation}
where $\ffdeg(s)>0$ is defined for each set $s$ of internal
lines\footnote{Note that the sets can be not connected.} of $G$
except the empty set and the set of all internal lines of $G$. The
probability density function is defined by
\begin{equation}\label{eq_density}
g(z_1,\ldots,z_n) =
\frac{g_0(z_1,\ldots,z_n)}{\int_{z_1,\ldots,z_n>0}
g_0(z_1,\ldots,z_n)\delta(z_1+\ldots+z_n-1) dz_1\ldots dz_n}.
\end{equation}
A fast random samples generation algorithm for a given $\ffdeg(s)$
is described in ~\cite{volkov_prd}.

Let us describe the procedure of obtaining $\ffdeg(s)$. The
following auxiliary definitions repeat the ones from
~\cite{volkov_prd}. By definition, put
$$
\omega(s)=2N_L(s)+|\electrons(s)|/2-|s|,
$$
where $|x|$ is the cardinality of a set $x$, $\electrons(s)$ is the
set of all electron lines in $s$, $N_L(s)$ is the number of
independent loops in $s$. If $s$ is the set of all internal lines of
a subgraph of $G$, then $\omega(s)$ coincides with the ultraviolet
degree of divergence of this subgraph that is defined above.

The problem of constructing a good $g(\underline{z})$ is very close
to the problem of obtaining a simple and close enough upper bound
for $|I(\underline{z})|$ and proving the integral finiteness, see
~\cite{volkov_prd}. Feynman-parametric expressions for the
integrands (without subtraction terms) can be represented as
fractions with denominators that vanish on the boundary of the
integration area, if we are on the mass shell ~\cite{volkov_2015}.
If we consider the nominators only, we can use the ultraviolet
degrees of divergence themselves, see ~\cite{speer}. If we take into
account the denominators too, the degrees must be increased, it is
performed by I-closures that are defined below. In addition to
vanishing denominators, the divergence subtraction complicates the
problem. The construction described below is based on both
theoretical considerations\footnote{Some of the ideas underlying the
concept of I-closure and this procedure of obtaining $\ffdeg(s)$
will be described in future papers (these ideas are quite
complicated and are not completely substantiated mathematically at
this moment).} and numerical experiments.

By $\iclos(s)$ we denote the set $s\cup s'$, where $s'$ is the set
of all internal photon lines $l$ in $G$ such that $s$ contains the
electron path in $G$ connecting the ends of $l$. The set $\iclos(s)$
is called the \emph{I-closure} of the set $s$.

By definition, put
$$
\omega'(s)=\omega(\iclos(s)).
$$

A graph $G''$ belonging to a forest $F\in\forests[G]$ is called a
\emph{child} of a graph $G'\in F$ in $F$ if $G''\varsubsetneq G'$,
and there is no $G'''\in F$ such that $G'''\varsubsetneq G'$,
$G''\varsubsetneq G'''$.

If $F\in\forests[G]$ and $G'\in F$ then by $G'/ F$ we denote the
graph that is obtained from $G'$ by shrinking all childs of $G'$ in
$F$ to points.

We also will use the symbols $\omega$, $\omega'$ for graphs $G'$
that are constructed from $G$ by some operations like described
above\footnote{See the corresponding examples in
~\cite{volkov_prd}.} and for sets $s$ that are subsets of the set of
internal lines of the whole graph $G$. We will denote it by
$\omega_{G'}(s)$ and $\omega'_{G'}(s)$, respectively. This means
that we apply the operations $\omega$ and $\omega'$ in the graph
$G'$ to the set $s'$ that is the intersection of $s$ and the set of
all internal lines of $G'$.

Electron self-energy subgraphs and lines joining them form chains
$l_1 G_1 l_2 G_2 \ldots l_r G_r l_{r+1}$, where $l_j$ are electron
lines of $G$, $G_j$ are electron self-energy subgraphs of $G$.
Maximal (with respect to inclusion) subsets
$\{l_1,l_2,\ldots,l_{r+1}\}$ corresponding to such chains are called
\emph{SE-chains}. The set of all SE-chains of $G$ is denoted by
$\sechains[G]$.

Suppose a graph $G'$ is constructed from $G$ by operations like
described above; by definition, put
$$
\omega^*_{G'}(s)=\omega'_{G'}(s)+\frac{1}{2}\sum_{\substack{s'\in\sechains[G]
\\ s'\subseteq s,\ s'\text{ in }G'}}(|s'|-1)
$$
(it is important that here we consider the SE-chains of the whole
graph $G$).

By $\forestsmax[G]$ we denote the set of all maximal forests
belonging to $\forests[G]$ (with respect to inclusion).

Let $\cbigf>0$, $\cbigz>0$, $\cadd$, $\csubi$, $\csubse$, $\csubo$
be constants. By definition, put
$$
\ffdeg(s) = \begin{cases}
\cbigz+\frac{(\cbigf-\cbigz)N_L(s)}{N_L(G)},\text{ if $s$ contain
all electron lines of $G$,} \\ \cadd + \min_{F\in\forestsmax[G]}
\sum_{G'\in F} \max(0, -\omega^*_{G'/ F}(s)-\ffsub[G']), \\ \quad
\text{otherwise,}\end{cases}
$$
where
$$
\ffsub[G']=\begin{cases}\csubi,\text{ if $G'\in\infragr[G]$,}
\\ \csubse,\text{ if $G'$ is a self-energy subgraph,} \\ \csubo\text{ in the other cases.}\end{cases}
$$
This formula for $\ffdeg(s)$ differs from the one that was defined
in ~\cite{volkov_prd} and gives better Monte Carlo convergence, if
appropriate values for constants are taken. For good Monte Carlo
convergence we can use the values
\begin{equation}\label{eq_mc_constants}
\begin{array}{c}
\cbigz=0.256,\ \cbigf=0.839,\ \cadd=0.786,\ \\  \csubi=0.2,\
\csubse=0 ,\ \csubo=0.2.
\end{array}
\end{equation}
These values were obtained by a series of numerical experiments on
4-loop Feynman graphs. See the examples for the considered
combinatorial constructions in ~\cite{volkov_prd}.

\section{REALIZATION AND NUMERICAL RESULTS}\label{sec_real}

\subsection{Overview}\label{subsec_real_overview}

\begin{figure}[H]
\begin{center}
\includegraphics[scale=1.0]{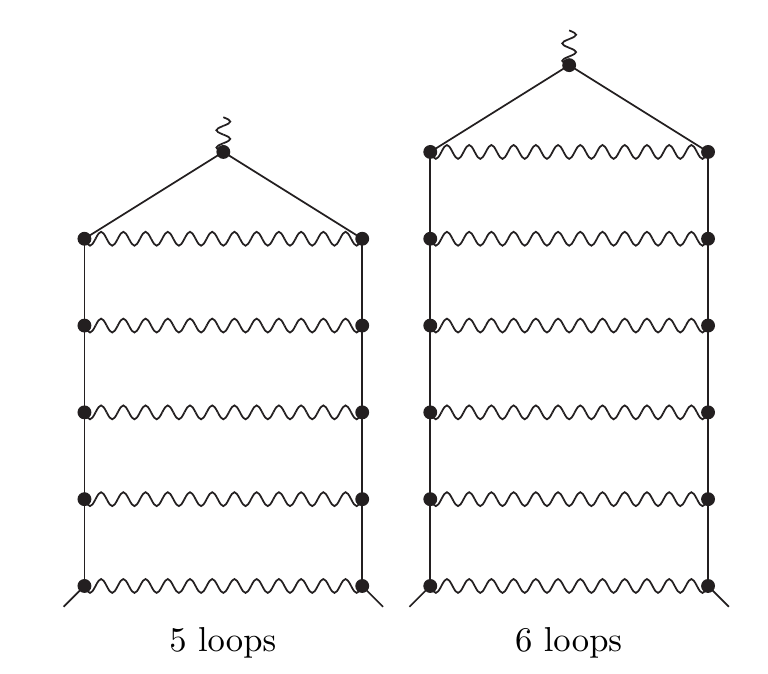}
\caption{5-loop and 6-loop ladder graphs} \label{fig_ladder}
\end{center}
\end{figure}

The computation on one GPU of NVidia Tesla K80 that was leased from
Google Cloud\footnote{using the free trial} showed the following
results ($1\sigma$ limits\footnote{See Section
\ref{subsec_error_est}}):
$$
A_1^{(4)}[\text{no lepton loops}]=-0.3441651(34),
$$
$$
 A_1^{(6)}[\text{no lepton
loops}]=0.90485(10),
$$
$$
A_1^{(8)}[\text{no lepton loops}]=-2.181(10),
$$
the corresponding computation times\footnote{d=days, h=hours,
min=minutes} are 21 h 37 min, 5 d 8 h, 7 d. The obtained
contributions of the 5-loop and 6-loop ladder graphs from FIG
\ref{fig_ladder} are $11.6530(58)$ and $34.31(20)$ respectively. The
corresponding computation times are 4 h 38 min and 8 h 24 min. All
obtained results are in good agreement with the known analytical and
semianalytical ones, see Table \ref{table_tech}. See also the
detailed results in Sections \ref{subsec_indi},
\ref{subsec_classes}, \ref{subsec_tech}.

We reduce the number of integration variables by one using the fact
that each integrand $I(z_1,\ldots,z_n)$ depends linearly on $z_a$
when $z_a+z_b$ is fixed, where $a$ and $b$ are the electron lines
that are incident to the vertex that is incident to the external
photon line, see ~\cite{volkov_prd}\footnote{and also
~\cite{kinoshita_rules}}. In contrast to
~\cite{volkov_2015,volkov_prd}, we use a nonadaptive\footnote{except
the selection of the parameters (\ref{eq_mc_constants}) and an
inter-graph adaptivity: the numbers of Monte Carlo samples for each
Feynman graph are selected to make the convergence maximally fast}
Monte Carlo algorithm. The absence of adaptivity simplifies a
realization on GPU and allows us to undertake an analysis of the
Monte Carlo samples behavior, see Section \ref{subsec_error_est}.

The D programming language ~\cite{dlang} was used for the generator
of the integrands code. The integrands and the Monte Carlo
integrator were written in C++ with CUDA ~\cite{cuda}. The integrand
code sizes are presented in Table \ref{table_tech}. The pseudorandom
generator MRG32k3a from the CURAND library ~\cite{curand} was used
for the Monte Carlo integration.

The integrand values are evaluated first using
double-precision\footnote{double-precision = 64 bit} floating point
operations that are fully supported on the GPU. If the
double-precision operations do not give enough accuracy, the
calculations are repeated using arbitrary-precision floating point
operations with increasing precision, see the details in Section
\ref{subsec_prec}.

All the integrand code is divided into shared libraries that are
linked dynamically with the integrator. Each Feynman graph and type
of arithmetic corresponds to one or several shared libraries. Each
of these shared libraries contains CUDA kernels\footnote{CUDA kernel
is a function in a program that is executed many times in parallel
on GPU and is called from the CPU part, see ~\cite{cuda}.} and
functions for calling them. For reducing the compilation
time\footnote{GPU device code is compiled very slowly, and the
compilation time increases rapidly with the size of functions.}
without losing the computation performance the size of the integrand
CUDA kernels is set at approximately 5000 operations. Also, for
reducing the compilation time each arbitrary-precision shared
library contains no more than 10 CUDA kernels.

The memory speed is a weak spot of GPU computing. So, the integrand
GPU code is organized in such a way that the most of the operations
are performed with the GPU register memory: we are trying to
minimize the number of the used variables, often to the detriment of
the arithmetic optimization.

To use the GPU parallel computing effectively we divide the Monte
Carlo samples for one Feynman graph into portions. Each portion
contains from $10^6$ to $10^8$ samples. First, we generate the
samples of a given portion and calculate the corresponding integrand
values in the fastest precision. After that, the samples requiring
an increased precision are collected and calculated. Each CUDA
kernel is launched on GPU in 19968 parallel threads\footnote{104
blocks of 192 threads}. To reduce the impact of the latency of CUDA
kernel calling each thread performs approximately 15 samples
sequentially in a loop.

\subsection{Interval arithmetic}\label{subsec_ia}

The interval arithmetic is an easy and reliable way for controlling
round-off errors. In this way all calculations are performed with
intervals, not with numbers. Arithmetic operations on intervals are
defined in such a way that each exact intermediate value $x$ is
quaranteed to be in the corresponding interval $[x^-;x^+]$. One can
use the following definitions:
$$
[x^-;x^+] + [y^-;y^+] = [(x^- + y^-)\rndd;(x^+ + y^+)\rndu],
$$
$$
[x^-;x^+] - [y^-;y^+] = [(x^- - y^+)\rndd;(x^+ - y^-)\rndu],
$$
$$
[x^-;x^+] \cdot [y^-;y^+] = [\min((x^-y^-)\rndd,(x^-y^+)\rndd,
(x^+y^-)\rndd, (x^+y^+)\rndd);$$ $$
\max((x^-y^-)\rndu,(x^-y^+)\rndu, (x^+y^-)\rndu, (x^+y^+)\rndu)],
$$
$$
1/[x^-;x^+] =
[\min((1/x^-)\rndd,(1/x^+)\rndd);\max((1/x^-)\rndu,(1/x^+)\rndu)],
$$
where $(*)\rndu$ and $(*)\rndd$ means the operation $(*)$ with
rounding up (to $+\infty$) or down (to $-\infty$). The most of
modern GPUs\footnote{as well as CPUs} support specifying rounding
mode for arithmetic operations and working with infinities for
handling overflows. Addition, subtraction and multiplication can be
realized directly by using the formulas proposed
above\footnote{Also, these formulas will work correctly with Not A
Numbers (NANs) despite the fact that the NVidia realization of
$\min$ and $\max$ ignores NANs in the lists of arguments.}. However,
for division it is required to perform additional operations for
handling divisions by zero and overflows. This does not slow down
the computation because the amount of divisions in the integrand
constructions is very small.

\subsection{Elimination of an interval arithmetic}\label{subsec_eia}

The direct interval arithmetic is a very slow thing. However, there
are many ways of increasing speed by weakening the distinctness of
the intervals.

We will use the following specificity of the integrands
construction. It is known ~\cite{volkov_2015}\footnote{See also
~\cite{zavialov,smirnov,kinoshita_rules}.} how to construct the
integrand for a given graph $G$ from the building blocks $V^{G'}$,
$Q^{G'}_{a,j}$, $B^{G'}_{ab}$, $S^{G'}$, where $G'$ is a graph that
can be obtained from a subgraph of $G$ by shrinking some subgraphs
to points, $a,b$ are internal electron lines of $G'$, $j=1,2$,
$V^{G'}$ is defined through a sum over 1-trees of $G'$,
$Q^{G'}_{a,j}$ through a sum over 1-trees\footnote{More precisely,
paper ~\cite{volkov_2015} has a definition of $\hat{Q}^{G'}_a$, the
values $Q^{G'}_{a,j}$ can be defined by
$\hat{Q}^{G'}_a=Q^{G'}_{a,1}\hat{p}_1+Q^{G'}_{a,2}\hat{p}_2$ in
terms of paper ~\cite{volkov_2015}.} passing $a$, $B^{G'}_{ab}$
through a sum over trees with cycle passing $a,b$, $S^{G'}$ through
a sum over 2-trees. See the full definitions in ~\cite{volkov_2015}.
The construction rules described in ~\cite{volkov_2015} give us to
observe that for a high number of independent loops in $G$ the most
part of the integrand computation is the calculations of polynomials
with the variables $Q^{G'}_{a,j}/V^{G'}$ and $B^{G'}_{ab}/V^{G'}$.

Suppose we want to calculate a polynomial of the intervals
$[x_1^-;x_1^+],\ldots,[x_n^-;x_n^+]$ that is constructed as a
sequence of additions, subtractions and multiplications. The main
ideas of the interval arithmetic elimination are:
\begin{itemize}
\item we can calculate the center of the resulting interval in
the direct double-precision arithmetic using the same polynomial
applied to the centers of $[x_j^-;x_j^+]$;
\item the radius of the resulting interval can be estimated as a
function of $x_j^-$, $x_j^+$ that is much more simple than the
source polynomial.
\end{itemize}

We will use the following inequality about the machine
double-precision arithmetic\footnote{The last term corresponds to
the case when a very small number is converted into zero after
rounding.}:
$$
|x-x\rnd|\leq 2^{-52}|x|+2^{-1022},
$$
where $x\rnd$ corresponds to the machine representation of $x$
rounded in any direction.

Let $x_j$ be the exact values corresponding to the intervals
$[x_j^-;x_j^+]$, $j=1,\ldots,n$. By $x_{n+1},\ldots,x_l$ we denote
the exact intermediate values that are obtained sequentially when we
calculate the value of the needed polynomial. To each $j=1,\ldots,n$
we assign a \emph{type} $t_j$: $t_j=0$ if $x_j$ is
$Q^{G'}_{a,k}/V^{G'}$, $t_j=1$ if $x_j$ is $B^{G'}_{ab}/V^{G'}$ (we
divide all source values into two groups in such a way because
$|Q^{G'}_{a,k}/V^{G'}|\leq 1$, but $B^{G'}_{ab}/V^{G'}$ are
unbounded\footnote{Generally speaking, we can divide them in any way
into any number of pieces. This splitting is selected as a
compromise between precision and speed.}). Let us define the numbers
$x_j\appr$, $M_j$, $\varepsilon_j$, $j=1,\ldots,l$, satisfying the
following conditions for all $j$:
\begin{itemize}
\item $|x_j\appr-x_j|\leq\varepsilon_j$;
\item $|x_j\appr|\leq M_j$.
\end{itemize}
We define them by using the following rules:
\begin{itemize}
\item $x_j\appr=((x_j^-+x_j^+)/2)\rnd$, $j=1,\ldots,n$ (thus,
$x_j\appr$ are the centers of the corresponding intervals; the
machine double-precision arithmetic guarantees that we always have
$x_j^-\leq x_j\appr\leq x_j^+$ if an overflow does not occur);
\item $M_j$ are defined for $j=1,\ldots,n$ by
$$
M_j=\max_{t_k=t_j} |x_k\appr|;
$$
\item $\varepsilon_j$ are defined for $j=1,\ldots,n$ by
$$
\varepsilon_j=\varepsilon=\max_{1\leq k\leq
n}\max\left((x_k\appr-x_k^+)\rndu, (x_k^+-x_k\appr)\rndu\right),
$$
\item if $x_j$ is obtained as $x_{k}*x_{r}$, where $*$ is the
addition, subtraction or multiplication, $j=n+1,\ldots,l$, then
$x_j\appr=(x_{k}\appr*x_{r}\appr)\rnd$ (thus, $x_j\appr$ are
obtained by the direct double-precision arithmetic without
specifying the rounding mode\footnote{In some tests, specifying a
rounding mode for addition or multiplication slows down the
performance of these operations on NVidia Tesla K80 by 7 times.
However, in the considering calculations it was not experienced, see
Table \ref{table_tech}.});
\item analogously, $(M_j,\varepsilon_j)$ is defined by
$$
(M_j,\varepsilon_j)=((M_k+M_r)(1+2^{-52})+2^{-1022},
\varepsilon_k+\varepsilon_r+2^{-52}(M_k+M_r)+2^{-1022})
$$
for addition and subtraction, and by
$$
(M_j,\varepsilon_j)=(M_k M_r(1+2^{-52})+2^{-1022}, $$
$$\varepsilon_k \varepsilon_r + \varepsilon_k M_r + \varepsilon_r
M_k + 2^{-52}M_k M_r+2^{-1022})
$$
for multiplication.
\end{itemize}

It is easy to see that for the final $l$ the value $\varepsilon_l$
can be expressed as a polynomial $P(M_{t=0},M_{t=1},\varepsilon)$
with positive coefficients in only three variables, where
$$
M_{t=a}=\max_{t_k=a} |x_k\appr|.
$$
Thus, the value of $\varepsilon_l$ can be obtained directly using
the coefficients of this polynomial without calculating the
intermediate values $M_k,\varepsilon_k$.

However, the polynomial
$$
P(M_{t=0},M_{t=1},\varepsilon)=\sum_{u,v,w}C_{u,v,w} (M_{t=0})^u
(M_{t=1})^v \varepsilon^w
$$
can still have many coefficients and therefore can require a lot of
arithmetic operations for computation. We estimate $P$ by another
expression in the following way. Let us split $P$ into four parts
$P_0$, $P_1$, $P_2$, $P_3$ by the following rules:
$$
P_0:\ C_{u,v,w}<2^{-100},\quad P_1:\ 2^{-100}\leq C_{u,v,w}<0.5,
$$
$$
P_2:\ C_{u,v,w}\geq 0.5,\ w\leq 1,\quad P_3:\ C_{u,v,w}\geq 0.5,\
w\geq 2.
$$
Thus, $P=P_0+P_1+P_2+P_3$. By definition, put
$$
u_j^-=\min_{C_{u,v,w}^j>0} u,\quad u_j^+=\max_{C_{u,v,w}^j>0} u,
$$
where
$$
P_j(M_{t=0},M_{t=1},\varepsilon)=\sum_{u,v,w}C_{u,v,w}^j (M_{t=0})^u
(M_{t=1})^v \varepsilon^w,\quad j=0,1,2,3.
$$
Let us define $v_j^-$, $v_j^+$, $w_j^-$, $w_j^+$ in analogous way.
Put
$$
P_j'(M_{t=0},M_{t=1},\varepsilon)=\left(\sum_{u,v,w}
C_{u,v,w}^j\right)
$$
$$
\cdot \max\left((M_{t=0})^{u_j^+},(M_{t=0})^{u_j^-}\right) \cdot
\max\left((M_{t=1})^{v_j^+},(M_{t=1})^{v_j^-}\right) \cdot
\max\left(\varepsilon^{w_j^+},\varepsilon^{w_j^-}\right).
$$
It is obvious that $P_j'\geq P_j$. So, we can use
$P'=P_0'+P_1'+P_2'+P_3'$ as a radius of the final interval, if it is
calculated by machine arithmetic operations with rounding
up\footnote{The coefficients $C_{u,v,w}$ and the sum of them must be
calculated with rounding up too. However, this calculation is
performed at the stage of codegeneration.}. $P'$ is much simpler for
calculation than $P$. Thus, an interval for the final value may
be\footnote{Overflows, infinities and NANs do not require an
additional consideration at all stages of the calculation.}
$$
[(x_l\appr-(P')\rndu)\rndd;(x_l\appr+(P')\rndu)\rndu].
$$
We split $P$ into four polynomials in such a way guiding the
following considerations:
\begin{itemize}
\item $P_3$ contains the most of the coefficients sum; however, its
contribution in $P_3'$ will be compensated by the multiplier
$\varepsilon^2$ (when $\varepsilon$ is near zero);
\item $P_2$ has a big sum of coefficients too; however, it is much
less than $P_3$ has; this sum will be compensated by the multiplier
$\varepsilon$ in $P_2'$;
\item $P_1$ has a little sum of coefficients; however, in some cases
$P_1'$ can be noticeable; thus, we separate $P_1$ from $P_0$ to
minimize the contribution of the $\max\cdot\max\cdot\max$ part in
the definition of $P_1'$;
\item the contribution of the coefficients of $P_0$ is always small.
\end{itemize}

\subsection{Algorithm of obtaining accurate integrand values}\label{subsec_prec}

We obtain the value\footnote{We can't use the double precision
directly for the probability density $g(\underline{z})$ because its
value sometimes goes beyond the range of double precision values.
This situation often occurs in the 6-loop case. We use the
representation $x\cdot 2^j$ instead, where the double precision is
used for $0.5\leq x<1$, the number $j$ is a 32-bit integer.}
$I(\underline{z})/g(\underline{z})$ from (\ref{eq_integral_appr})
first by the eliminated interval arithmetic from Section
\ref{subsec_eia}. If the obtained interval $[y^-;y^+]$ does not
satisfy the condition $y^+-y^-\leq \sigma/4$, where $\sigma$ is the
current error estimation\footnote{In the beginning of the integral
computation we calculate from $10^5$ to $10^7$ points in the direct
double-precision interval arithmetic taking the nearest to zero
points for each interval.} for the obtained integral value, we
repeat the calculation in the direct double-precision interval
arithmetic. If it is not enough, we reiterate this calculation in
the interval arithmetic based on floating-point numbers with 128-bit
mantissa and with 256-bit mantissa (if needed). If the 256-bit
mantissa precision is not enough, we suppose that the value equals
$0$.

The arithmetic with 128-bit mantissa is realized on GPU in such a
way that all operations are performed with the GPU register memory.
The arithmetic with 256-bit mantissa works with the global GPU
memory. The usage of the register memory improves the performance by
about 10 times\footnote{However, Table \ref{table_tech} shows a gap
that is much more than 10 times. The reason is that there are very
few points requiring 256-bit mantissa, we can't use GPU parallelism
effectively.}.

We also use a routine for prevention of emerging occasional very
large values that is analogous to the one described in
~\cite{volkov_prd}, but adapted for GPU parallel computing.

\subsection{Modified probability density functions}

It is theoretically possible the situation when $g(\underline{z})$
from (\ref{eq_integral_appr}) is very small, but the smallness of
$|I(\underline{z})|$ does not correspond to it. An emergence of such
situations can make the Monte Carlo convergence worse. For patching
it we use the probability density functions
$$
g(\underline{z})=C_1g_1(\underline{z}) + C_2g_2(\underline{z}) +
C_3g_3(\underline{z}) + C_4g_4(\underline{z})
$$
instead of (\ref{eq_density_0}), (\ref{eq_density}), where $g_1$ is
defined by (\ref{eq_density_0}), (\ref{eq_density}),
$$
g_2(z_1,\ldots,z_n)=\frac{\prod_{l=2}^n \left[
\ffdeg(\{j_l,j_{l+1},\ldots,j_n\})(z_{j_l}/z_{j_{l-1}})^{\ffdeg(\{j_l,j_{l+1},\ldots,j_n\})}
\right]}{n! z_1z_2\ldots z_n},
$$
when the definitions from Section \ref{sec_monte_carlo} are used,
$g_3$ is defined by (\ref{eq_density_0}), (\ref{eq_density}), but
with same $\ffdeg(s)=D$, $g_4(\underline{z})=(n-1)!$ (the uniform
distribution). To generate a random sample with the distribution
$g(\underline{z})$ we should perform the following two steps:
\begin{itemize}
\item generate randomly $j=1,2,3,4$, where the probability of
selecting $j$ is $C_j$;
\item generate a sample with the distribution $g_j(\underline{z})$.
\end{itemize}
The generation with the distribition $g_2(\underline{z})$ is the
same as for distributions defined by (\ref{eq_density_0}),
(\ref{eq_density}), but at the stage of sector generation we must
take sectors with same probabilities, see ~\cite{volkov_prd}. All
computations are performed with the following values for the
constants:
$$
D=0.75,\ C_2=0.03,\ C_3=0.035,\ C_4=0.035,\ C_1=1-C_2-C_3-C_4.
$$

\subsection{Monte Carlo error estimation}\label{subsec_error_est}

Let $\underline{z}_1,\ldots,\underline{z}_N$ be random samples, the
formula (\ref{eq_integral_appr}) is used for Monte Carlo
integration. By definition, put
$y_j=I(\underline{z})/g(\underline{z})$. The conventional error
estimation approach is based on the following formula for the
standard deviation:
$$
(\sigma_{\downarrow})^2=\frac{\sum_{j=1}^N y_j^2}{N^2} -
\frac{\left(\sum_{j=1}^N y_j \right)^2}{N^3}.
$$
However, this formula has a tendency to underestimate the real
standard deviation. Let us consider the 5-loop and 6-loop ladder
examples. By definition, put
$$
\maxlog = \max_j \lfloor \log_2 |y_j| + 0.5 \rfloor,
$$
let $n_k$ be the quantity of samples $j$ such that
\begin{equation}\label{eq_n_prob_def}
2^{\maxlog-k-0.5}\leq |y_j| < 2^{\maxlog-k+0.5}.
\end{equation}
$\maxlog$ and $n_j$ for the 5-loop and 6-loop ladders are presented
in Table \ref{table_distr_ladder}. $n_k$ is an approximation for
$Np_k$, where $p_k$ is the probability that a sample is in the
interval (\ref{eq_n_prob_def}). We can see that the real standard
deviation is highly dependent on the behavior of $p_j$ for $j<0$.
For example, if $p_{j+1}/p_j<4$ for all $j<j_0$ then the standard
deviation is infinite\footnote{Table \ref{table_distr_ladder}
demonstrates that for the 6-loop ladder such a situation is quite
possible.}.

We will use the improved estimation\footnote{When we calculate
deviation probabilities based on the standard deviation we use a
presupposition based on the Central Limit Theorem that the
distribution of $\sum_{j=1}^N y_j/N$ is close to the Gauss normal
distribution. However, it is difficult to estimate the difference
between the real distribution and the normal one. For example, the
Berry-Esseen inequality uses the third central moment of random
variables that is infinite if $p_{j+1}/p_j<8$ for all $j<j_0$ (Table
\ref{table_distr_ladder} shows that this situation is quite possible
for both 5-loop and 6-loop ladders).}
$$
(\sigma_{\uparrow})^2=(\sigma_{\downarrow})^2+\triangle\uncert +
\triangle\peak,
$$
where\footnote{The definitions of $\sigma_{\downarrow}$ and
$\triangle\uncert$ repeat the ones from ~\cite{volkov_prd}.}
$$
\triangle\uncert = 4\cdot \max_{k=0}^{19} 4^{\maxlog-k}\sqrt{n_k}
$$
is the contribution of the uncertainty of $n_k$, $\triangle\peak$ is
the contribution of the predicted behavior of $p_j$ for $j<0$ that
is described below\footnote{This procedure is a result of tests on
different graph contributions to $a_e$. It is developed for future
calculations of contributions to $a_e$ of higher orders. It should
not be treated as a universal procedure that works for all Monte
Carlo integrations. However, a big value of
$\sigma_{\uparrow}/\sigma_{\downarrow}$ indicates that the obtained
error estimation is unreliable.}.

The idea is to approximate $n_j$ by a geometric progression taking
into account that $n_j$ are known with an uncertainty of about
$C\sqrt{n_j}$ and that $p_{j+1}/p_j$ changes with $j$.

Put
$$
h_j=\begin{cases} \log_2 n_j, \text{ if } n_j>0, \\ -2, \text{ if }
n_j=0, \end{cases}
$$
$$
h_j^{\pm} = \log_2\max\left(\frac{1}{8}, n_j+\frac{1}{2}\pm
\sqrt{n_j+\frac{1}{4}} \right),
$$
Here $h_j$ is an approximated value of $\log_2(Np_j)$,
$[h_j^-;h_j^+]$ is an interval for this value that is obtained
taking into account that $n_j$ is known with
uncertainty\footnote{$x=n+\frac{C^2}{2} \pm C\sqrt{n+\frac{C^2}{4}}$
is the solution of the equation $x\mp C\sqrt{x}=n$.}.

We will estimate the absolute value of a difference between
neighbour $\log_2(p_{j+1}/p_j)$ by the value $d$, where
$$
d=\max_{0\leq j<k\leq 18} \frac{d_{jk}}{k-j},
$$
where $d_{jk}$ is the distance from $0$ to the interval
$[d_{jk}^-;d_{jk}^+]$,
$$
d_{jk}^- = (h_{k+1}^- - h_k^+) - (h_{j+1}^+ - h_j^-),\quad d_{jk}^+
= (h_{k+1}^+ - h_k^-) - (h_{j+1}^- - h_j^+).
$$

For approximation of the sequence by a progression we will use
another values for $\log_2(Np_j)$ uncertainty that are obtained
taking into account that errors for lesser $j$ are more critical:
$$
u_j=\begin{cases} \frac{1}{2} \left[ \log_2\left(n_j +
\frac{C_j^2}{2} + C_j\sqrt{n_j + \frac{C_j^2}{4}}\right) -
\log_2\left(n_j + \frac{C_j^2}{2} - C_j\sqrt{n_j +
\frac{C_j^2}{4}}\right)\right], \\ \quad \text{ if }n_j > 0, \\ 3,
\text{ if } n_j=0,\end{cases}
$$
where
$$
C_j = \frac{2}{1 + \frac{2(j+1)}{20}}.
$$
For approximating the sequence of logarithms by a linear function
$kj+b$ let us introduce coefficients $a^l_j$, $f^l_j$, $2\leq l\leq
20$, $0\leq j<l$, for the least squares method\footnote{The explitit
formulas are $a^l_j=\frac{12j-6(l-1)}{l(l^2-1)}$,
$f^l_j=\frac{2(2l-1)-6j}{l(l+1)}$.}:
$$
\left( \sum_{j=0}^{l-1} a^l_j x_j; \sum_{j=0}^{l-1} f^l_j x_j
\right) = \argmin_{(k;b)} \sum_{j=0}^{l-1}(kj+b-x_j)^2
$$
for all $l$ and $x_0,\ldots,x_{l-1}$.

Put
$$
k_l=\sum_{j=0}^{l-1} a^l_jh_j - \sqrt{\sum_{j=0}^{l-1}(a^l_j)^2
u_j^2} - d\sum_{j=0}^{l-1}\frac{j(j-1)a^l_j}{2}, $$
$$
k=\max(k_2,\ldots,k_{20},h_0-1-u_0).
$$
This formula takes into account both uncertainty of $n_j$ and shift
of $p_{j+1}/p_j$ with $j$. We take $\max$ to prevent from excessive
overestimation\footnote{The last argument of $\max$ is needed to
process the situation when $n_0$ is quite big: in this case an
absence of $n_{-1}$ is very informative.}. Also, put
$$
\triangle b = \min_l \left( \sqrt{\sum_{j=0}^{l-1} (f^l_j)^2 u_j^2 }
+ d \sum_{j=0}^{l-1}\frac{j(j-1)f^l_j}{2} \right),
$$
$b=\sum_{j=0}^{l-1} f^l_j h_j$, where we take $l$ for which the
minimum is achieved. Let us define $\triangle\peak$ by
$$
\triangle\peak=2^{2\cdot\maxlog + b + 0.7\triangle
b}\left(\frac{1}{1-2^{-w}}-1\right),
$$
where
$$
w=\frac{k-\frac{17}{8}+\sqrt{\left(k-\frac{17}{8}\right)^2 +
\frac{1}{16}}}{2}+\frac{1}{8}.
$$
The meaning of this definition is that we use the formula for the
sum of a geometric progression taking $w$ instead of $k-2$. $w$ is
defined in such a way that $w\sim k-2$ as $k\rightarrow +\infty$ and
$w\rightarrow 1/8$ as $k\rightarrow -\infty$.

We use $\sigma_{\uparrow}$ for all numerical results that are
presented in this paper.

Tables \ref{table_dependence_4loops}, \ref{table_dependence_5loops},
\ref{table_dependence_6loops} contain the dependence of the error
estimations and the real errors on numbers of samples
$N_{\text{total}}$ for $A_1^{(8)}[\text{no lepton loops}]$, 5-loop
and 6-loop ladders respectively.

\begin{longtable}{ccc}\caption{Probability distributions for 5-loop ladder and 6-loop ladder (continued)} \\
\hline \hline
 Parameter & 5-loop ladder & 6-loop ladder\\ \hline  \endhead
\caption{Probability distributions for 5-loop ladder and 6-loop ladder}\label{table_distr_ladder} \\
\hline \hline
 Parameter & 5-loop ladder & 6-loop ladder\\ \hline  \endfirsthead
\hline \endfoot  \hline \hline \endlastfoot
\verb"maxlog" & $23$ & $28$ \\
$n_{0}$ & $11$ & $2$ \\
$n_{1}$ & $64$ & $8$ \\
$n_{2}$ & $393$ & $45$ \\
$n_{3}$ & $2300$ & $174$ \\
$n_{4}$ & $11891$ & $785$ \\
$n_{5}$ & $51840$ & $2898$ \\
$n_{6}$ & $204817$ & $9374$ \\
$n_{7}$ & $688060$ & $25759$ \\
$n_{8}$ & $1885211$ & $62363$ \\
$n_{9}$ & $4300121$ & $135343$ \\
$n_{10}$ & $8615210$ & $267630$ \\
$n_{11}$ & $15701395$ & $490720$ \\
$n_{12}$ & $26582404$ & $849862$ \\
$n_{13}$ & $42456874$ & $1394740$ \\
$n_{14}$ & $64590501$ & $2198221$ \\
$n_{15}$ & $94011212$ & $3331999$ \\
$n_{16}$ & $131314678$ & $4892615$ \\
$n_{17}$ & $176228467$ & $6965326$ \\
$n_{18}$ & $228021742$ & $9626392$ \\
$n_{19}$ & $285614048$ & $12965533$ \\
\end{longtable}
\small
\begin{longtable}{cccccc}\caption{Dependence of the estimated error and the difference between the obtained value and the known semianalytical one ~\cite{laporta_8}
on the number of Monte Carlo samples $N_{\text{total}}$:
$A_1^{(8)}[\text{no lepton loops}]$,
see a remark about $\sigma_{\uparrow}$, $\sigma_{\downarrow}$ calculation in Section \ref{subsec_classes} (continued)} \\
\hline \hline $N_{\text{total}}$ & Value & $\sigma_{\uparrow}$ &
$\sigma_{\downarrow}$ & Difference &
$\sigma_{\uparrow}/\sigma_{\downarrow}$ \\ \hline  \endhead
\caption{Dependence of the estimated error and the difference
between the obtained value and the known semianalytical one
~\cite{laporta_8} on the number of Monte Carlo samples
$N_{\text{total}}$: $A_1^{(8)}[\text{no lepton loops}]$,
see a remark about $\sigma_{\uparrow}$, $\sigma_{\downarrow}$ calculation in Section \ref{subsec_classes} }\label{table_dependence_4loops} \\
\hline \hline $N_{\text{total}}$ & Value & $\sigma_{\uparrow}$ &
$\sigma_{\downarrow}$ & Difference &
$\sigma_{\uparrow}/\sigma_{\downarrow}$ \\ \hline  \endfirsthead
\hline \endfoot  \hline \hline \endlastfoot
$40\times 10^{9}$ & $-2.3937$ & $0.2144$ & $0.1168$ & $-0.2168$ & $1.84$ \\
$10^{11}$ & $-2.2323$ & $0.0710$ & $0.0494$ & $-0.0555$ & $1.44$ \\
$20\times 10^{10}$ & $-2.1820$ & $0.0468$ & $0.0345$ & $-0.0051$ & $1.36$ \\
$50\times 10^{10}$ & $-2.1851$ & $0.0282$ & $0.0218$ & $-0.0083$ & $1.30$ \\
$10^{12}$ & $-2.1757$ & $0.0194$ & $0.0154$ & $0.0012$ & $1.26$ \\
$20\times 10^{11}$ & $-2.1702$ & $0.0133$ & $0.0109$ & $0.0066$ & $1.23$ \\
$32\times 10^{11}$ & $-2.1807$ & $0.0104$ & $0.0086$ & $-0.0038$ & $1.21$ \\
\end{longtable}
\normalsize \small
\begin{longtable}{cccccc}\caption{Dependence of the estimated error and the difference between the obtained value and the known analytical one ~\cite{caffo} on the number of Monte Carlo samples $N_{\text{total}}$: 5-loop ladder (continued)} \\
\hline \hline $N_{\text{total}}$ & Value & $\sigma_{\uparrow}$ &
$\sigma_{\downarrow}$ & Difference &
$\sigma_{\uparrow}/\sigma_{\downarrow}$ \\ \hline  \endhead
\caption{Dependence of the estimated error and the difference between the obtained value and the known analytical one ~\cite{caffo} on the number of Monte Carlo samples $N_{\text{total}}$: 5-loop ladder}\label{table_dependence_5loops} \\
\hline \hline $N_{\text{total}}$ & Value & $\sigma_{\uparrow}$ &
$\sigma_{\downarrow}$ & Difference &
$\sigma_{\uparrow}/\sigma_{\downarrow}$ \\ \hline  \endfirsthead
\hline \endfoot  \hline \hline \endlastfoot
$59\times 10^{5}$ & $12.0682$ & $0.8202$ & $0.3288$ & $0.4090$ & $2.49$ \\
$12\times 10^{7}$ & $11.6120$ & $0.1349$ & $0.0720$ & $-0.0472$ & $1.87$ \\
$24\times 10^{7}$ & $11.6934$ & $0.0800$ & $0.0525$ & $0.0342$ & $1.52$ \\
$60\times 10^{7}$ & $11.6798$ & $0.0665$ & $0.0379$ & $0.0206$ & $1.76$ \\
$10^{9}$ & $11.6678$ & $0.0427$ & $0.0270$ & $0.0086$ & $1.58$ \\
$20\times 10^{8}$ & $11.6474$ & $0.0277$ & $0.0192$ & $-0.0118$ & $1.44$ \\
$50\times 10^{8}$ & $11.6448$ & $0.0150$ & $0.0120$ & $-0.0144$ & $1.25$ \\
$10^{10}$ & $11.6509$ & $0.0111$ & $0.0086$ & $-0.0083$ & $1.29$ \\
$20\times 10^{9}$ & $11.6541$ & $0.0073$ & $0.0061$ & $-0.0051$ & $1.19$ \\
$29\times 10^{9}$ & $11.6530$ & $0.0058$ & $0.0050$ & $-0.0062$ & $1.16$ \\
\end{longtable}
\normalsize \small
\begin{longtable}{cccccc}\caption{Dependence of the estimated error and the difference between the obtained value and the known analytical one ~\cite{caffo} on the number of Monte Carlo samples $N_{\text{total}}$: 6-loop ladder (continued)} \\
\hline \hline $N_{\text{total}}$ & Value & $\sigma_{\uparrow}$ &
$\sigma_{\downarrow}$ & Difference &
$\sigma_{\uparrow}/\sigma_{\downarrow}$ \\ \hline  \endhead
\caption{Dependence of the estimated error and the difference between the obtained value and the known analytical one ~\cite{caffo} on the number of Monte Carlo samples $N_{\text{total}}$: 6-loop ladder}\label{table_dependence_6loops} \\
\hline \hline $N_{\text{total}}$ & Value & $\sigma_{\uparrow}$ &
$\sigma_{\downarrow}$ & Difference &
$\sigma_{\uparrow}/\sigma_{\downarrow}$ \\ \hline  \endfirsthead
\hline \endfoot  \hline \hline \endlastfoot
$15\times 10^{6}$ & $34.3209$ & $7.1538$ & $2.0690$ & $-0.0461$ & $3.46$ \\
$65\times 10^{7}$ & $35.4566$ & $1.1201$ & $0.4659$ & $1.0896$ & $2.40$ \\
$97\times 10^{7}$ & $35.0500$ & $0.7556$ & $0.3566$ & $0.6829$ & $2.12$ \\
$12\times 10^{8}$ & $35.0187$ & $0.6808$ & $0.3201$ & $0.6517$ & $2.13$ \\
$22\times 10^{8}$ & $34.5855$ & $0.4217$ & $0.2276$ & $0.2185$ & $1.85$ \\
$41\times 10^{8}$ & $34.3967$ & $0.3020$ & $0.1675$ & $0.0297$ & $1.80$ \\
$70\times 10^{8}$ & $34.3651$ & $0.2320$ & $0.1337$ & $-0.0019$ & $1.74$ \\
$10^{10}$ & $34.3062$ & $0.1974$ & $0.1137$ & $-0.0608$ & $1.74$ \\
\end{longtable}
\normalsize

\subsection{Contributions of individual Feynman graphs}\label{subsec_indi}

\begin{figure}[H]
\begin{center}
\includegraphics[scale=0.5]{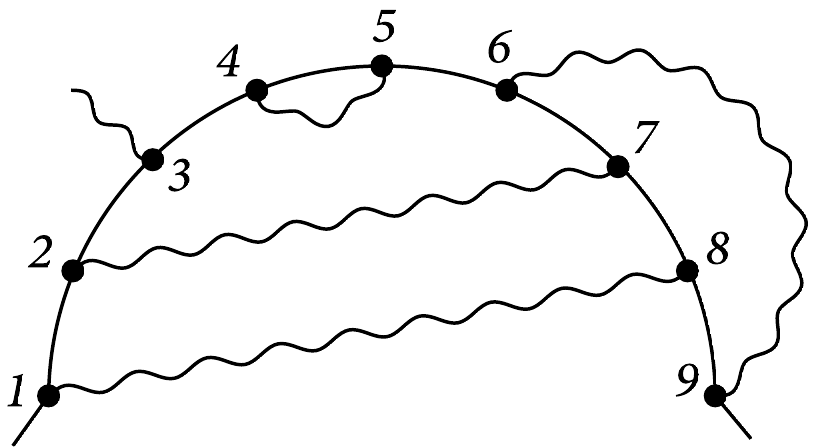}
\caption{4-loop Feynman graph: example} \label{fig_code_example}
\end{center}
\end{figure}

\begin{figure}[H]
\begin{center}
\includegraphics[scale=1.0]{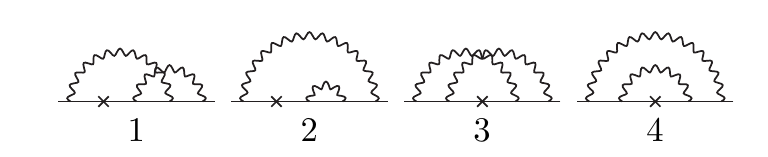}
\caption{2-loop Feynman graphs without lepton loops}
\label{fig_2loops}
\end{center}
\end{figure}

\begin{figure}[H]
\begin{center}
\includegraphics[scale=1.0]{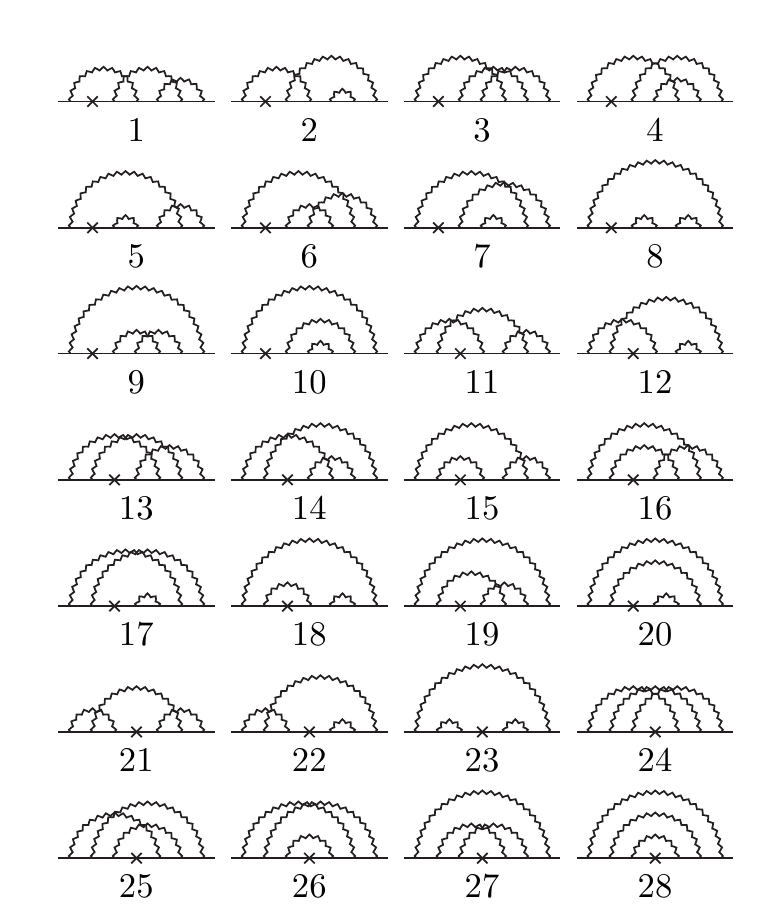}
\caption{3-loop Feynman graphs without lepton loops}
\label{fig_3loops}
\end{center}
\end{figure}

The contributions of 2-loop and 3-loop Feynman graphs to $A_1^{(4)}$
and $A_1^{(6)}$ are presented in Tables \ref{table_indi_2loops} and
\ref{table_indi_3loops}. The corresponding pictures are FIG.
\ref{fig_2loops} and FIG. \ref{fig_3loops}. Each individual
contribution in this paper is given for a Feynman graph without
arrow directions on electron lines and includes the contributions of
the corresponding graphs with all directions (that are the same).
The 4-loop graphs are split into gauge invariant classes $(k,m,m')$,
where $m$ and $m'$ are numbers of internal photon lines to the left
and to the right from the external photon line (or vice versa), $k$
is the number of photons with the ends on the opposite sides of it.
We do not give a picture for 4-loop graphs, but they are encoded in
the tables as expressions of the form
$$
p;\ s_1-f_1,\ s_2-f-2,\ s_3-f_3,\ s_4-f_4,
$$
where $p$ is the number of vertex that is incident to the external
photon line, $s_j$ and $f_j$ are the ends of the $j$-th internal
photon line, the vertexes are enumerated from $1$ to $9$ along the
electron path, $s_j<f_j$, $s_1<\ldots<s_4$. The graphs are ordered
lexicographically, and we guarantee that the code of a graph is the
lexicographically minimal one. For example, the code of the graph
from FIG. \ref{fig_code_example} is
$$
3;\ 1-8,\ 2-7,\ 4-5,\ 6-9.
$$
The contributions of the 4-loop graphs are presented in Tables
\ref{table_indi_4loops_1_3_0}, \ref{table_indi_4loops_2_2_0},
\ref{table_indi_4loops_1_2_1}, \ref{table_indi_4loops_3_1_0},
\ref{table_indi_4loops_2_1_1}, \ref{table_indi_4loops_4_0_0}. The
numbers of the graphs for which the contribution must coincide with
the contribution obtained by the direct subtraction on the mass
shell in Feynman gauge are marked by a star $^*$, see Section
\ref{subsec_classes}.

The fields of the tables have the following meaning:
\begin{itemize}
\item Value is the obtained value for the contribution
 with the uncertainty $\sigma_{\uparrow}$, see Section
\ref{subsec_error_est};
\item $\sigma_{\uparrow}/\sigma_{\downarrow}$ is the relation
between the improved standard deviation and the conventional one,
see Section \ref{subsec_error_est};
\item $N_{\text{total}}$ is the total quantity of Monte Carlo samples;
\item $N^{\text{fail}}_{\text{EIA}}$ is the quantity of samples for
which the Eliminated Interval Arithmetic from Section
\ref{subsec_eia} failed;
\item $\triangle^{\text{fail}}_{\text{EIA}}$ is the contribution of
that samples\footnote{Sometimes this contribution can be many times
more than the total 4-loop contribution! See, for example, graph 157
from Table \ref{table_indi_4loops_1_2_1}. However, the Eliminated
Interval Arithmetic significantly improves the computation
performance, see Table \ref{table_tech}.};
\item $N^{\text{fail}}_{\text{IA}}$ is the quantity of samples for
which the direct double-precision interval arithmetic from Section
\ref{subsec_ia} failed;
\item $\triangle^{\text{fail}}_{\text{IA}}$ is the contribution of
that samples;
\item $N^{\text{fail}}_{\text{128}}$ is the quantity of samples for
which the interval arithmetic based on numbers with 128-bit mantissa
failed;
\item $\triangle^{\text{fail}}_{\text{128}}$ is the contribution of
that samples\footnote{Even these contributions can be noticeable.
See, for example, graph 134 from Table
\ref{table_indi_4loops_2_2_0}.}.
\end{itemize}
\tiny
\begin{longtable}{ccccccccccc}\caption{Contributions of individual Feynman graphs from FIG. \ref{fig_2loops} to $A^{(4)}_1$ (continued)} \\
\hline \hline Number & Graph & Value & $N_{\text{total}}$ &
$N_{\text{EIA}}^{\text{fail}}$ & $N_{\text{IA}}^{\text{fail}}$ &
$N_{\text{128}}^{\text{fail}}$ &
$\triangle_{\text{EIA}}^{\text{fail}}$ &
$\triangle_{\text{IA}}^{\text{fail}}$ &
$\triangle_{\text{128}}^{\text{fail}}$ &
$\sigma_{\uparrow}/\sigma_{\downarrow}$ \\ \hline  \endhead
\caption{Contributions of individual Feynman graphs from FIG. \ref{fig_2loops} to $A^{(4)}_1$}\label{table_indi_2loops} \\
\hline \hline Number & Graph & Value & $N_{\text{total}}$ &
$N_{\text{EIA}}^{\text{fail}}$ & $N_{\text{IA}}^{\text{fail}}$ &
$N_{\text{128}}^{\text{fail}}$ &
$\triangle_{\text{EIA}}^{\text{fail}}$ &
$\triangle_{\text{IA}}^{\text{fail}}$ &
$\triangle_{\text{128}}^{\text{fail}}$ &
$\sigma_{\uparrow}/\sigma_{\downarrow}$ \\ \hline  \endfirsthead
\hline \endfoot  \hline \hline \endlastfoot
$1$ & 2;\,1-4,\,3-5 & -0.0640193(19) & $94\times 10^{10}$ & $26\times 10^{8}$ & $32\times 10^{4}$ & $0$ & $0.003$ & $2\times 10^{-6}$ & $0$ & 1.04 \\
$2$ & 2;\,1-5,\,3-4 & -0.5899758(14) & $58\times 10^{10}$ & $61\times 10^{7}$ & $50\times 10^{6}$ & $2$ & $-0.0005$ & $-2\times 10^{-6}$ & $-2\times 10^{-19}$ & 1.00 \\
$3$ & 3;\,1-4,\,2-5 & -0.4676475(17) & $90\times 10^{10}$ & $44\times 10^{7}$ & $57479$ & $0$ & $-0.008$ & $-10^{-5}$ & $0$ & 1.05 \\
$4$ & 3;\,1-5,\,2-4 & 0.7774774(18) & $92\times 10^{10}$ & $34\times 10^{8}$ & $17\times 10^{6}$ & $0$ & $0.007$ & $0.0002$ & $0$ & 1.00 \\
\end{longtable}
\normalsize

\tiny
\begin{longtable}{ccccccccccc}\caption{Contributions of individual Feynman graphs from FIG. \ref{fig_3loops} to $A^{(6)}_1$ (continued)} \\
\hline \hline Number & Graph & Value & $N_{\text{total}}$ &
$N_{\text{EIA}}^{\text{fail}}$ & $N_{\text{IA}}^{\text{fail}}$ &
$N_{\text{128}}^{\text{fail}}$ &
$\triangle_{\text{EIA}}^{\text{fail}}$ &
$\triangle_{\text{IA}}^{\text{fail}}$ &
$\triangle_{\text{128}}^{\text{fail}}$ &
$\sigma_{\uparrow}/\sigma_{\downarrow}$ \\ \hline  \endhead
\caption{Contributions of individual Feynman graphs from FIG. \ref{fig_3loops} to $A^{(6)}_1$}\label{table_indi_3loops} \\
\hline \hline Number & Graph & Value & $N_{\text{total}}$ &
$N_{\text{EIA}}^{\text{fail}}$ & $N_{\text{IA}}^{\text{fail}}$ &
$N_{\text{128}}^{\text{fail}}$ &
$\triangle_{\text{EIA}}^{\text{fail}}$ &
$\triangle_{\text{IA}}^{\text{fail}}$ &
$\triangle_{\text{128}}^{\text{fail}}$ &
$\sigma_{\uparrow}/\sigma_{\downarrow}$ \\ \hline  \endfirsthead
\hline \endfoot  \hline \hline \endlastfoot
$1$ & 2;\,1-4,\,3-6,\,5-7 & -1.679616(20) & $29\times 10^{10}$ & $57\times 10^{8}$ & $33\times 10^{4}$ & $0$ & $-0.1$ & $-5\times 10^{-5}$ & $0$ & 1.08 \\
$2$ & 2;\,1-4,\,3-7,\,5-6 & 0.832792(20) & $28\times 10^{10}$ & $40\times 10^{8}$ & $26\times 10^{6}$ & $0$ & $0.1$ & $0.0009$ & $0$ & 1.10 \\
$3$ & 2;\,1-5,\,3-6,\,4-7 & 0.214875(14) & $19\times 10^{10}$ & $22\times 10^{8}$ & $88\times 10^{4}$ & $1$ & $0.01$ & $7\times 10^{-5}$ & $-4\times 10^{-24}$ & 1.05 \\
$4$ & 2;\,1-5,\,3-7,\,4-6 & -0.028928(11) & $11\times 10^{10}$ & $30\times 10^{8}$ & $10^{6}$ & $2$ & $-0.004$ & $-3\times 10^{-5}$ & $-10^{-11}$ & 1.03 \\
$5$ & 2;\,1-6,\,3-4,\,5-7 & -0.097163(26) & $47\times 10^{10}$ & $16\times 10^{9}$ & $13\times 10^{7}$ & $10$ & $-0.002$ & $10^{-5}$ & $3\times 10^{-16}$ & 1.16 \\
$6$ & 2;\,1-6,\,3-5,\,4-7 & 0.144471(12) & $14\times 10^{10}$ & $31\times 10^{8}$ & $23\times 10^{4}$ & $3$ & $0.06$ & $9\times 10^{-6}$ & $-2\times 10^{-36}$ & 1.02 \\
$7$ & 2;\,1-6,\,3-7,\,4-5 & 0.804106(17) & $22\times 10^{10}$ & $26\times 10^{8}$ & $18\times 10^{6}$ & $1$ & $0.02$ & $-0.0001$ & $-10^{-61}$ & 1.08 \\
$8$ & 2;\,1-7,\,3-4,\,5-6 & -2.123267(16) & $17\times 10^{10}$ & $34\times 10^{8}$ & $48\times 10^{7}$ & $4496$ & $-0.02$ & $-0.0002$ & $-10^{-12}$ & 1.00 \\
\end{longtable}
\normalsize

\tiny
\begin{longtable}{ccccccccccc}\caption{Contributions of graphs from the gauge-invariant class $(1,3,0)$ to $A^{(8)}_1$ (continued)} \\
\hline \hline Number & Graph & Value & $N_{\text{total}}$ &
$N_{\text{EIA}}^{\text{fail}}$ & $N_{\text{IA}}^{\text{fail}}$ &
$N_{\text{128}}^{\text{fail}}$ &
$\triangle_{\text{EIA}}^{\text{fail}}$ &
$\triangle_{\text{IA}}^{\text{fail}}$ &
$\triangle_{\text{128}}^{\text{fail}}$ &
$\sigma_{\uparrow}/\sigma_{\downarrow}$ \\ \hline  \endhead
\caption{Contributions of graphs from the gauge-invariant class $(1,3,0)$ to $A^{(8)}_1$}\label{table_indi_4loops_1_3_0} \\
\hline \hline Number & Graph & Value & $N_{\text{total}}$ &
$N_{\text{EIA}}^{\text{fail}}$ & $N_{\text{IA}}^{\text{fail}}$ &
$N_{\text{128}}^{\text{fail}}$ &
$\triangle_{\text{EIA}}^{\text{fail}}$ &
$\triangle_{\text{IA}}^{\text{fail}}$ &
$\triangle_{\text{128}}^{\text{fail}}$ &
$\sigma_{\uparrow}/\sigma_{\downarrow}$ \\ \hline  \endfirsthead
\hline \endfoot  \hline \hline \endlastfoot
$1$ & 2;\,1-4,\,3-6,\,5-8,\,7-9 & 2.19701(73) & $15\times 10^{9}$ & $92\times 10^{7}$ & $27780$ & $0$ & $0.6$ & $0.0001$ & $0$ & 1.56 \\
$2$ & 2;\,1-4,\,3-6,\,5-9,\,7-8 & -3.81327(91) & $23\times 10^{9}$ & $95\times 10^{7}$ & $11\times 10^{5}$ & $1$ & $-1$ & $-0.01$ & $2\times 10^{-45}$ & 1.65 \\
$3$ & 2;\,1-4,\,3-7,\,5-8,\,6-9 & 0.55330(31) & $49\times 10^{8}$ & $23\times 10^{7}$ & $31677$ & $1$ & $0.2$ & $0.0001$ & $9\times 10^{-29}$ & 1.52 \\
$4$ & 2;\,1-4,\,3-7,\,5-9,\,6-8 & 1.82177(56) & $10^{10}$ & $57\times 10^{7}$ & $47919$ & $2$ & $0.5$ & $0.0002$ & $10^{-36}$ & 1.51 \\
$5$ & 2;\,1-4,\,3-8,\,5-6,\,7-9 & -2.43257(93) & $23\times 10^{9}$ & $13\times 10^{8}$ & $29\times 10^{5}$ & $1$ & $-0.8$ & $-0.008$ & $10^{-35}$ & 1.72 \\
$6$ & 2;\,1-4,\,3-8,\,5-7,\,6-9 & 0.95204(51) & $89\times 10^{8}$ & $50\times 10^{7}$ & $26741$ & $1$ & $0.3$ & $8\times 10^{-5}$ & $-4\times 10^{-55}$ & 1.41 \\
$7$ & 2;\,1-4,\,3-8,\,5-9,\,6-7 & -2.19745(69) & $15\times 10^{9}$ & $57\times 10^{7}$ & $54\times 10^{4}$ & $0$ & $-0.4$ & $-0.003$ & $0$ & 1.89 \\
$8$ & 2;\,1-4,\,3-9,\,5-6,\,7-8 & 2.1481(10) & $20\times 10^{9}$ & $75\times 10^{7}$ & $27\times 10^{6}$ & $161$ & $0.8$ & $0.03$ & $-6\times 10^{-7}$ & 1.65 \\
\end{longtable}
\normalsize

\tiny
\begin{longtable}{ccccccccccc}\caption{Contributions of graphs from the gauge-invariant class $(2,2,0)$ to $A^{(8)}_1$ (continued)} \\
\hline \hline Number & Graph & Value & $N_{\text{total}}$ &
$N_{\text{EIA}}^{\text{fail}}$ & $N_{\text{IA}}^{\text{fail}}$ &
$N_{\text{128}}^{\text{fail}}$ &
$\triangle_{\text{EIA}}^{\text{fail}}$ &
$\triangle_{\text{IA}}^{\text{fail}}$ &
$\triangle_{\text{128}}^{\text{fail}}$ &
$\sigma_{\uparrow}/\sigma_{\downarrow}$ \\ \hline  \endhead
\caption{Contributions of graphs from the gauge-invariant class $(2,2,0)$ to $A^{(8)}_1$}\label{table_indi_4loops_2_2_0} \\
\hline \hline Number & Graph & Value & $N_{\text{total}}$ &
$N_{\text{EIA}}^{\text{fail}}$ & $N_{\text{IA}}^{\text{fail}}$ &
$N_{\text{128}}^{\text{fail}}$ &
$\triangle_{\text{EIA}}^{\text{fail}}$ &
$\triangle_{\text{IA}}^{\text{fail}}$ &
$\triangle_{\text{128}}^{\text{fail}}$ &
$\sigma_{\uparrow}/\sigma_{\downarrow}$ \\ \hline  \endfirsthead
\hline \endfoot  \hline \hline \endlastfoot
$75$ & 3;\,1-4,\,2-6,\,5-8,\,7-9 & -10.44260(93) & $25\times 10^{9}$ & $13\times 10^{8}$ & $11\times 10^{4}$ & $5$ & $-3$ & $-0.003$ & $-3\times 10^{-32}$ & 1.29 \\
$76$ & 3;\,1-4,\,2-6,\,5-9,\,7-8 & 10.0730(12) & $43\times 10^{9}$ & $17\times 10^{8}$ & $23\times 10^{5}$ & $3$ & $3$ & $0.08$ & $2\times 10^{-6}$ & 1.72 \\
$77$ & 3;\,1-4,\,2-7,\,5-8,\,6-9 & -1.67666(34) & $56\times 10^{8}$ & $27\times 10^{7}$ & $55710$ & $2$ & $-0.6$ & $-0.0003$ & $-10^{-12}$ & 1.32 \\
$78$ & 3;\,1-4,\,2-7,\,5-9,\,6-8 & -5.75797(76) & $18\times 10^{9}$ & $99\times 10^{7}$ & $12\times 10^{4}$ & $3$ & $-2$ & $-0.003$ & $-6\times 10^{-24}$ & 1.31 \\
$79$ & 3;\,1-4,\,2-8,\,5-6,\,7-9 & 11.5103(11) & $31\times 10^{9}$ & $20\times 10^{8}$ & $42\times 10^{5}$ & $1$ & $5$ & $0.2$ & $4\times 10^{-24}$ & 1.32 \\
$80$ & 3;\,1-4,\,2-8,\,5-7,\,6-9 & -5.15144(69) & $15\times 10^{9}$ & $77\times 10^{7}$ & $79826$ & $0$ & $-2$ & $-0.003$ & $0$ & 1.12 \\
$81$ & 3;\,1-4,\,2-8,\,5-9,\,6-7 & 6.80288(85) & $23\times 10^{9}$ & $85\times 10^{7}$ & $84\times 10^{4}$ & $3$ & $2$ & $0.04$ & $-4\times 10^{-38}$ & 1.63 \\
$82$ & 3;\,1-4,\,2-9,\,5-6,\,7-8 & -10.3320(13) & $31\times 10^{9}$ & $15\times 10^{8}$ & $47\times 10^{6}$ & $207$ & $-4$ & $-0.6$ & $-8\times 10^{-5}$ & 1.45 \\
\end{longtable}
\normalsize

\tiny
\begin{longtable}{ccccccccccc}\caption{Contributions of graphs from the gauge-invariant class $(1,2,1)$ to $A^{(8)}_1$ (continued)} \\
\hline \hline Number & Graph & Value & $N_{\text{total}}$ &
$N_{\text{EIA}}^{\text{fail}}$ & $N_{\text{IA}}^{\text{fail}}$ &
$N_{\text{128}}^{\text{fail}}$ &
$\triangle_{\text{EIA}}^{\text{fail}}$ &
$\triangle_{\text{IA}}^{\text{fail}}$ &
$\triangle_{\text{128}}^{\text{fail}}$ &
$\sigma_{\uparrow}/\sigma_{\downarrow}$ \\ \hline  \endhead
\caption{Contributions of graphs from the gauge-invariant class $(1,2,1)$ to $A^{(8)}_1$}\label{table_indi_4loops_1_2_1} \\
\hline \hline Number & Graph & Value & $N_{\text{total}}$ &
$N_{\text{EIA}}^{\text{fail}}$ & $N_{\text{IA}}^{\text{fail}}$ &
$N_{\text{128}}^{\text{fail}}$ &
$\triangle_{\text{EIA}}^{\text{fail}}$ &
$\triangle_{\text{IA}}^{\text{fail}}$ &
$\triangle_{\text{128}}^{\text{fail}}$ &
$\sigma_{\uparrow}/\sigma_{\downarrow}$ \\ \hline  \endfirsthead
\hline \endfoot  \hline \hline \endlastfoot
$149$ & 4;\,1-3,\,2-6,\,5-8,\,7-9 & 13.6554(10) & $27\times 10^{9}$ & $18\times 10^{8}$ & $75542$ & $4$ & $5$ & $0.01$ & $-2\times 10^{-45}$ & 1.08 \\
$150$ & 4;\,1-3,\,2-6,\,5-9,\,7-8 & -12.6376(13) & $41\times 10^{9}$ & $22\times 10^{8}$ & $43\times 10^{5}$ & $3$ & $-5$ & $-0.2$ & $-10^{-56}$ & 1.32 \\
$151$ & 4;\,1-3,\,2-7,\,5-8,\,6-9 & 2.72526(52) & $99\times 10^{8}$ & $64\times 10^{7}$ & $10^{5}$ & $4$ & $1$ & $0.0008$ & $5\times 10^{-48}$ & 1.50 \\
$152$ & 4;\,1-3,\,2-7,\,5-9,\,6-8 & 6.70242(77) & $17\times 10^{9}$ & $11\times 10^{8}$ & $11\times 10^{4}$ & $3$ & $3$ & $0.01$ & $10^{-26}$ & 1.04 \\
$153$ & 4;\,1-3,\,2-8,\,5-6,\,7-9 & -15.7206(11) & $26\times 10^{9}$ & $23\times 10^{8}$ & $11\times 10^{6}$ & $4$ & $-7$ & $-0.4$ & $2\times 10^{-8}$ & 1.02 \\
$154$ & 4;\,1-3,\,2-8,\,5-7,\,6-9 & 5.17997(78) & $17\times 10^{9}$ & $11\times 10^{8}$ & $93380$ & $2$ & $3$ & $0.01$ & $10^{-43}$ & 1.02 \\
$155$ & 4;\,1-3,\,2-8,\,5-9,\,6-7 & -9.33944(81) & $18\times 10^{9}$ & $97\times 10^{7}$ & $14\times 10^{5}$ & $4$ & $-3$ & $-0.1$ & $-10^{-49}$ & 1.05 \\
$156$ & 4;\,1-3,\,2-9,\,5-6,\,7-8 & 18.6188(12) & $20\times 10^{9}$ & $18\times 10^{8}$ & $60\times 10^{6}$ & $1162$ & $7$ & $0.9$ & $0.0001$ & 1.04 \\
\end{longtable}
\normalsize

\tiny
\begin{longtable}{ccccccccccc}\caption{Contributions of graphs from the gauge-invariant class $(3,1,0)$ to $A^{(8)}_1$ (continued)} \\
\hline \hline Number & Graph & Value & $N_{\text{total}}$ &
$N_{\text{EIA}}^{\text{fail}}$ & $N_{\text{IA}}^{\text{fail}}$ &
$N_{\text{128}}^{\text{fail}}$ &
$\triangle_{\text{EIA}}^{\text{fail}}$ &
$\triangle_{\text{IA}}^{\text{fail}}$ &
$\triangle_{\text{128}}^{\text{fail}}$ &
$\sigma_{\uparrow}/\sigma_{\downarrow}$ \\ \hline  \endhead
\caption{Contributions of graphs from the gauge-invariant class $(3,1,0)$ to $A^{(8)}_1$}\label{table_indi_4loops_3_1_0} \\
\hline \hline Number & Graph & Value & $N_{\text{total}}$ &
$N_{\text{EIA}}^{\text{fail}}$ & $N_{\text{IA}}^{\text{fail}}$ &
$N_{\text{128}}^{\text{fail}}$ &
$\triangle_{\text{EIA}}^{\text{fail}}$ &
$\triangle_{\text{IA}}^{\text{fail}}$ &
$\triangle_{\text{128}}^{\text{fail}}$ &
$\sigma_{\uparrow}/\sigma_{\downarrow}$ \\ \hline  \endfirsthead
\hline \endfoot  \hline \hline \endlastfoot
$169$ & 4;\,1-5,\,2-6,\,3-8,\,7-9 & -1.02160(39) & $67\times 10^{8}$ & $28\times 10^{7}$ & $29\times 10^{4}$ & $26$ & $-0.3$ & $0.002$ & $10^{-8}$ & 1.23 \\
$170$ & 4;\,1-5,\,2-6,\,3-9,\,7-8 & 0.82043(44) & $83\times 10^{8}$ & $26\times 10^{7}$ & $83\times 10^{4}$ & $64$ & $0.3$ & $0.004$ & $-2\times 10^{-7}$ & 1.54 \\
$171^*$ & 4;\,1-5,\,2-7,\,3-8,\,6-9 & -1.35615(40) & $88\times 10^{8}$ & $89\times 10^{6}$ & $14\times 10^{4}$ & $6$ & $-0.3$ & $-0.004$ & $10^{-10}$ & 1.39 \\
$172$ & 4;\,1-5,\,2-7,\,3-9,\,6-8 & -0.88139(29) & $42\times 10^{8}$ & $14\times 10^{7}$ & $56\times 10^{4}$ & $115$ & $-0.2$ & $0.009$ & $2\times 10^{-7}$ & 1.18 \\
$173$ & 4;\,1-5,\,2-8,\,3-6,\,7-9 & -4.37354(62) & $14\times 10^{9}$ & $61\times 10^{7}$ & $31\times 10^{4}$ & $3$ & $-2$ & $-0.0004$ & $7\times 10^{-24}$ & 1.26 \\
$174^*$ & 4;\,1-5,\,2-8,\,3-7,\,6-9 & 0.16235(32) & $59\times 10^{8}$ & $87\times 10^{6}$ & $21\times 10^{4}$ & $3$ & $0.1$ & $0.0009$ & $6\times 10^{-15}$ & 1.81 \\
$175$ & 4;\,1-5,\,2-8,\,3-9,\,6-7 & 0.91185(27) & $44\times 10^{8}$ & $79\times 10^{6}$ & $46\times 10^{4}$ & $40$ & $0.1$ & $-0.002$ & $-6\times 10^{-8}$ & 1.27 \\
$176$ & 4;\,1-5,\,2-9,\,3-6,\,7-8 & 4.01347(73) & $19\times 10^{9}$ & $72\times 10^{7}$ & $14\times 10^{5}$ & $2$ & $2$ & $0.05$ & $2\times 10^{-29}$ & 1.41 \\
\end{longtable}
\normalsize

\tiny
\begin{longtable}{ccccccccccc}\caption{Contributions of graphs from the gauge-invariant class $(2,1,1)$ to $A^{(8)}_1$ (continued)} \\
\hline \hline Number & Graph & Value & $N_{\text{total}}$ &
$N_{\text{EIA}}^{\text{fail}}$ & $N_{\text{IA}}^{\text{fail}}$ &
$N_{\text{128}}^{\text{fail}}$ &
$\triangle_{\text{EIA}}^{\text{fail}}$ &
$\triangle_{\text{IA}}^{\text{fail}}$ &
$\triangle_{\text{128}}^{\text{fail}}$ &
$\sigma_{\uparrow}/\sigma_{\downarrow}$ \\ \hline  \endhead
\caption{Contributions of graphs from the gauge-invariant class $(2,1,1)$ to $A^{(8)}_1$}\label{table_indi_4loops_2_1_1} \\
\hline \hline Number & Graph & Value & $N_{\text{total}}$ &
$N_{\text{EIA}}^{\text{fail}}$ & $N_{\text{IA}}^{\text{fail}}$ &
$N_{\text{128}}^{\text{fail}}$ &
$\triangle_{\text{EIA}}^{\text{fail}}$ &
$\triangle_{\text{IA}}^{\text{fail}}$ &
$\triangle_{\text{128}}^{\text{fail}}$ &
$\sigma_{\uparrow}/\sigma_{\downarrow}$ \\ \hline  \endfirsthead
\hline \endfoot  \hline \hline \endlastfoot
$223$ & 5;\,1-3,\,2-6,\,4-8,\,7-9 & -6.61670(58) & $11\times 10^{9}$ & $61\times 10^{7}$ & $58456$ & $0$ & $-2$ & $-0.002$ & $0$ & 1.08 \\
$224$ & 5;\,1-3,\,2-6,\,4-9,\,7-8 & 10.3187(10) & $30\times 10^{9}$ & $18\times 10^{8}$ & $37\times 10^{5}$ & $1$ & $4$ & $0.1$ & $4\times 10^{-66}$ & 1.53 \\
$225$ & 5;\,1-3,\,2-7,\,4-8,\,6-9 & 0.70044(49) & $95\times 10^{8}$ & $46\times 10^{7}$ & $32\times 10^{4}$ & $26$ & $0.06$ & $0.0007$ & $-10^{-7}$ & 1.52 \\
$226$ & 5;\,1-3,\,2-7,\,4-9,\,6-8 & -2.37520(44) & $73\times 10^{8}$ & $50\times 10^{7}$ & $10^{5}$ & $1$ & $-0.5$ & $-2\times 10^{-5}$ & $-7\times 10^{-44}$ & 1.16 \\
$227$ & 5;\,1-3,\,2-8,\,4-6,\,7-9 & 4.07903(69) & $13\times 10^{9}$ & $13\times 10^{8}$ & $52\times 10^{4}$ & $1$ & $1$ & $0.02$ & $-10^{-46}$ & 1.03 \\
$228$ & 5;\,1-3,\,2-8,\,4-7,\,6-9 & 2.09761(43) & $87\times 10^{8}$ & $38\times 10^{7}$ & $69528$ & $2$ & $0.7$ & $0.001$ & $10^{-40}$ & 1.30 \\
$229$ & 5;\,1-3,\,2-8,\,4-9,\,6-7 & 3.36347(78) & $19\times 10^{9}$ & $83\times 10^{7}$ & $10^{6}$ & $3$ & $0.5$ & $5\times 10^{-5}$ & $-2\times 10^{-17}$ & 1.38 \\
$230$ & 5;\,1-3,\,2-9,\,4-6,\,7-8 & -9.9012(11) & $24\times 10^{9}$ & $29\times 10^{8}$ & $20\times 10^{6}$ & $309$ & $-3$ & $-0.3$ & $-2\times 10^{-5}$ & 1.04 \\
\end{longtable}
\normalsize

\tiny
\begin{longtable}{ccccccccccc}\caption{Contributions of graphs from the gauge-invariant class $(4,0,0)$ to $A^{(8)}_1$ (continued)} \\
\hline \hline Number & Graph & Value & $N_{\text{total}}$ &
$N_{\text{EIA}}^{\text{fail}}$ & $N_{\text{IA}}^{\text{fail}}$ &
$N_{\text{128}}^{\text{fail}}$ &
$\triangle_{\text{EIA}}^{\text{fail}}$ &
$\triangle_{\text{IA}}^{\text{fail}}$ &
$\triangle_{\text{128}}^{\text{fail}}$ &
$\sigma_{\uparrow}/\sigma_{\downarrow}$ \\ \hline  \endhead
\caption{Contributions of graphs from the gauge-invariant class $(4,0,0)$ to $A^{(8)}_1$}\label{table_indi_4loops_4_0_0} \\
\hline \hline Number & Graph & Value & $N_{\text{total}}$ &
$N_{\text{EIA}}^{\text{fail}}$ & $N_{\text{IA}}^{\text{fail}}$ &
$N_{\text{128}}^{\text{fail}}$ &
$\triangle_{\text{EIA}}^{\text{fail}}$ &
$\triangle_{\text{IA}}^{\text{fail}}$ &
$\triangle_{\text{128}}^{\text{fail}}$ &
$\sigma_{\uparrow}/\sigma_{\downarrow}$ \\ \hline  \endfirsthead
\hline \endfoot  \hline \hline \endlastfoot
$253^*$ & 5;\,1-6,\,2-7,\,3-8,\,4-9 & 0.29657(24) & $49\times 10^{8}$ & $57\times 10^{6}$ & $22\times 10^{4}$ & $8$ & $0.2$ & $0.01$ & $3\times 10^{-10}$ & 1.36 \\
$254^*$ & 5;\,1-6,\,2-7,\,3-9,\,4-8 & -0.47196(32) & $55\times 10^{8}$ & $70\times 10^{6}$ & $64\times 10^{4}$ & $97$ & $-0.1$ & $-0.003$ & $-6\times 10^{-9}$ & 1.65 \\
$255^*$ & 5;\,1-6,\,2-8,\,3-7,\,4-9 & -0.57757(12) & $21\times 10^{8}$ & $31\times 10^{6}$ & $23\times 10^{4}$ & $31$ & $-0.1$ & $-0.001$ & $-2\times 10^{-7}$ & 1.42 \\
$256^*$ & 5;\,1-6,\,2-8,\,3-9,\,4-7 & 0.21265(21) & $39\times 10^{8}$ & $77\times 10^{6}$ & $10^{5}$ & $0$ & $-0.01$ & $-5\times 10^{-5}$ & $0$ & 1.62 \\
$257^*$ & 5;\,1-6,\,2-9,\,3-7,\,4-8 & -1.01853(40) & $79\times 10^{8}$ & $16\times 10^{7}$ & $40\times 10^{4}$ & $6$ & $-0.4$ & $-0.002$ & $-7\times 10^{-17}$ & 1.48 \\
$258^*$ & 5;\,1-6,\,2-9,\,3-8,\,4-7 & -0.01236(43) & $95\times 10^{8}$ & $27\times 10^{7}$ & $46\times 10^{4}$ & $12$ & $-0.1$ & $-0.006$ & $10^{-18}$ & 1.54 \\
$259^*$ & 5;\,1-7,\,2-6,\,3-9,\,4-8 & 0.49710(18) & $32\times 10^{8}$ & $29\times 10^{6}$ & $16\times 10^{4}$ & $3$ & $0.09$ & $0.0005$ & $8\times 10^{-22}$ & 1.40 \\
$260$ & 5;\,1-7,\,2-8,\,3-9,\,4-6 & 0.60670(24) & $43\times 10^{8}$ & $21\times 10^{7}$ & $17\times 10^{4}$ & $6$ & $0.1$ & $0.0008$ & $-2\times 10^{-27}$ & 1.23 \\
\end{longtable}
\normalsize

\subsection{Classes of Feynman graphs}\label{subsec_classes}

The contributions and $N_{\text{total}}$ for all classes in this
paper are obtained as sums of the corresponding individual values.
The values $\sigma_{\uparrow}$, $\sigma_{\downarrow}$ for the
classes are obtained by
$$
\sigma_{\uparrow}=\sqrt{\sum_j (\sigma_{\uparrow,j})^2},\quad
\sigma_{\downarrow}=\sqrt{\sum_j (\sigma_{\downarrow,j})^2},
$$
where $\sigma_{\uparrow,j}$ and $\sigma_{\downarrow,j}$ are the
corresponding individual values.

The contributions of graph sets to $A_1^{(4)}$, $A_1^{(6)}$,
$A_1^{(8)}$ for comparison with the direct subtraction on the mass
shell in the Feynman gauge are presented in Tables
\ref{table_split_2loops}, \ref{table_split_3loops},
\ref{table_split_4loops}. The 2-loop and 3-loop tables include a
comparison with the known analytical results\footnote{The big
discrepancy for the set $14,17$ in Table \ref{table_split_3loops} is
probably caused by an unstable behavior of the pseudorandom
generator MRG32k3a. The generator Philox\underline{ }4x32\underline{
}10 ~\cite{curand} seems to work better on this set.} and with the
old results from ~\cite{volkov_2015}\footnote{The uncertainties in
~\cite{volkov_2015} correspond to 90\% confidential limits (under
the assumption that the probability distribution is Gauss normal).}.
Table \ref{table_split_4loops} does not include one-element sets,
these sets (individual graphs) are marked by a star in the tables
containing individual contributions.

The contributions of the gauge invariant classes $(k,m,m')$ (see the
definition in Section \ref{subsec_indi}) and their comparison with
the semianalytical results from ~\cite{laporta_8} are presented in
Table \ref{table_gauge_4loops}.

The equivalence of the subtraction procedure from Section
\ref{sec_subtraction} and the direct subtraction on the mass shell
for all presented sets can be proved in a combinatorial
way\footnote{if we do not consider the matter of divergence
regularizations}. Let us consider an example: the set $26,27$ from
Table \ref{table_split_3loops}. The contribution of this set can be
schematically written as
\begin{figure}[H]
\begin{center}
\includegraphics[scale=0.5]{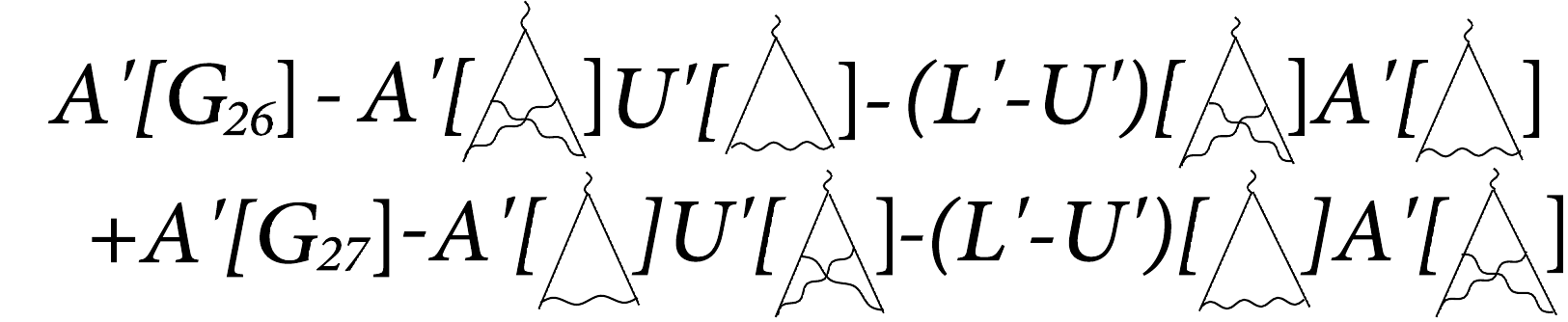}
\end{center}
\end{figure}
Here, $A'$, $L'$, $U'$ are operators that are applied to Feynman
amplitudes and return numbers,
$$
A\Gamma_{\mu}=e\gamma_{\mu}(A'\Gamma_{\mu}),\quad
L\Gamma_{\mu}=e\gamma_{\mu}(L'\Gamma_{\mu}), \quad
U\Gamma_{\mu}=e\gamma_{\mu}(U'\Gamma_{\mu}),
$$
the definitions from Section \ref{sec_subtraction} are used, a
constant multiplier is omitted. Analogously, the corresponding
contribution that is obtained by the direct subtraction on the mass
shell is
\begin{figure}[H]
\begin{center}
\includegraphics[scale=0.5]{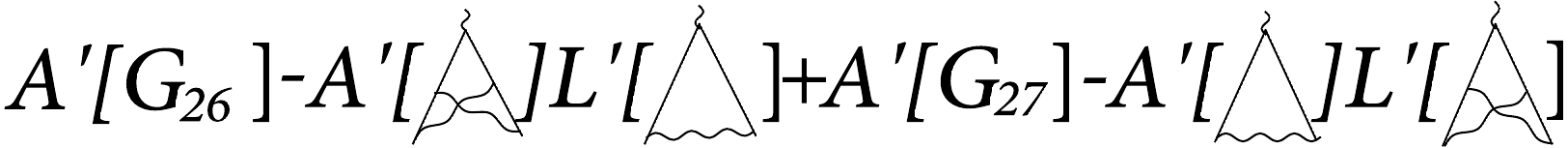}
\end{center}
\end{figure}
It is easy to see that these expressions are equivalent. Let us
consider another example: the set $11,17$ from Table
\ref{table_split_3loops}. The contribution of this set is
\begin{figure}[H]
\begin{center}
\includegraphics[scale=0.5]{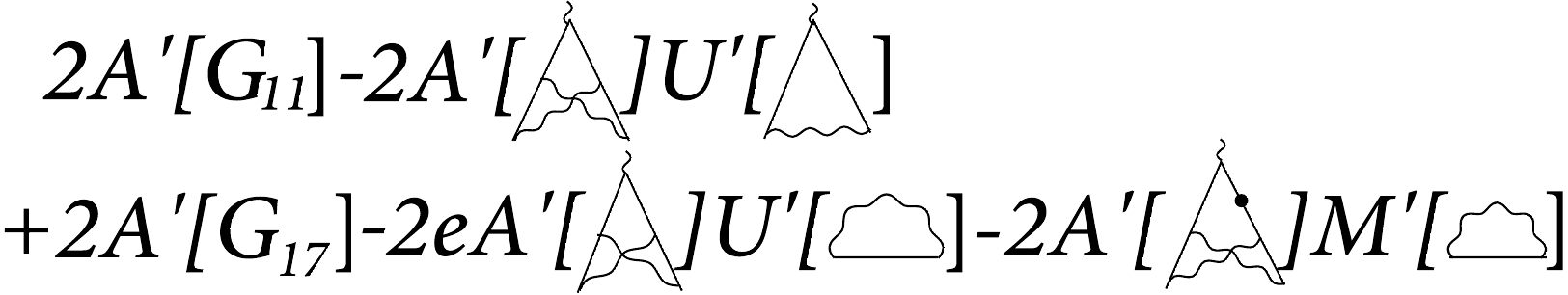}
\end{center}
\end{figure}
Here, the operators $U'$ and $M'$ that are applied to Feynman
amplitudes of self-energy subgraphs are defined by
$$
U\Sigma(p)=e[M'\Sigma+(U'\Sigma)(\hat{p}-m)].
$$
The terms containing $U'$ are cancelled because $U$ preserves the
Ward identity, see ~\cite{volkov_2015}. An analogous cancelation
works for the direct subtraction expression and leads to the same
result.

Sometimes for proving the equivalence it is needed to use the Ward
identity for individual Feynman graphs, see ~\cite{peskin}. For
example, for the operator $U'$ we can use the following equality:
\begin{figure}[H]
\begin{center}
\includegraphics[scale=0.5]{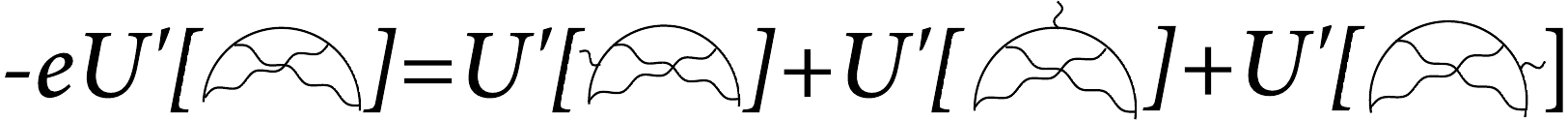}
\end{center}
\end{figure}
The right part of this equality contains all possible insertions of
an external photon line to the graph from the left part.

\small
\begin{longtable}{cccc}\caption{Contributions to $A^{(4)}_1$ (see FIG. \ref{fig_2loops}) that must coincide with the values that are obtained by direct subtraction on the mass shell in the Feynman gauge, a comparison of these results with the values from ~\cite{analyt2_p} and with the old values from ~\cite{volkov_2015} (continued)} \\
\hline \hline Set of graphs & Value & Analytical value & Value from
~\cite{volkov_2015} \\ \hline  \endhead
\caption{Contributions to $A^{(4)}_1$ (see FIG. \ref{fig_2loops}) that must coincide with the values that are obtained by direct subtraction on the mass shell in the Feynman gauge, a comparison of these results with the values from ~\cite{analyt2_p} and with the old values from ~\cite{volkov_2015}}\label{table_split_2loops} \\
\hline \hline Set of graphs & Value & Analytical value & Value from
~\cite{volkov_2015} \\ \hline  \endfirsthead \hline \endfoot  \hline
\hline \endlastfoot
1-2 & -0.6539950(23) & $-0.653998963627$ & $-0.654032(54)$ \\
3 & -0.4676475(17) & $-0.467645446094$ & $-0.467626(44)$ \\
4 & 0.7774774(18) & $0.777478022283$ & $0.777455(52)$ \\
\end{longtable}
\normalsize

\small
\begin{longtable}{ccccc}\caption{Contributions to $A^{(6)}_1$ (see FIG. \ref{fig_3loops}) that must coincide with the values that are obtained by direct subtraction on the mass shell in the Feynman gauge, a comparison of these results with the known analytical values and with the old values from ~\cite{volkov_2015} (continued)} \\
\hline \hline Set of graphs & Value & Analytical value & Reference &
Value from ~\cite{volkov_2015} \\ \hline  \endhead
\caption{Contributions to $A^{(6)}_1$ (see FIG. \ref{fig_3loops}) that must coincide with the values that are obtained by direct subtraction on the mass shell in the Feynman gauge, a comparison of these results with the known analytical values and with the old values from ~\cite{volkov_2015}}\label{table_split_3loops} \\
\hline \hline Set of graphs & Value & Analytical value &
Reference\footnotemark & Value from ~\cite{volkov_2015} \\ \hline
\endfirsthead \hline \endfoot  \hline \hline \endlastfoot
1-10\footnotetext{More precisely, the expressions from ~\cite{analyt_c} are semianalytical. The corresponding analytical expressions are given in ~\cite{laporta_1993}.} & 0.533289(54) & $0.533355$ & ~\cite{analyt_b,analyt_e,analyt_d,analyt_c,analyt_f,analyt3,analyt_h} & $0.5340(18)$ \\
11-12 & 1.541644(37) & $1.541649$ & ~\cite{analyt_e,analyt_c} & $1.5436(34)$ \\
13 & -1.757945(15) & $-1.757936$ & ~\cite{analyt3} & $-1.7579(10)$ \\
14, 17 & 0.455517(26) & $0.455452$ & ~\cite{analyt_f,analyt_h} & $0.4549(14)$ \\
15, 18-20 & -0.402749(46) & $-0.402717$ & ~\cite{analyt_b,analyt_e} & $-0.4030(41)$ \\
16 & -0.334691(14) & $-0.334695$ & ~\cite{analyt_f} & $-0.33468(95)$ \\
21-23 & 0.421080(43) & $0.421171$ & ~\cite{analyt_b,analyt_e,analyt_c} & $0.4207(22)$ \\
24 & -0.0267956(78) & $-0.026799$ & ~\cite{analyt3} & $-0.02688(47)$ \\
25 & 1.861914(17) & $1.861908$ & ~\cite{analyt_f} & $1.8629(14)$ \\
26-27 & -3.176700(22) & $-3.176685$ & ~\cite{analyt_d,analyt_h} & $-3.1764(22)$ \\
28 & 1.790285(19) & $1.790278$ & ~\cite{analyt_d} & $1.7888(19)$ \\
\end{longtable}
\normalsize

\footnotesize
\begin{longtable}{cccc}\caption{Contributions to $A^{(8)}_1$ that must coincide with the values that are obtained by direct subtraction on the mass shell in the Feynman gauge (continued)} \\
\hline \hline Set of graphs & Value & $N_{\text{total}}$ &
$\sigma_{\uparrow}/\sigma_{\downarrow}$ \\ \hline  \endhead
\caption{Contributions to $A^{(8)}_1$ that must coincide with the values that are obtained by direct subtraction on the mass shell in the Feynman gauge}\label{table_split_4loops} \\
\hline \hline Set of graphs & Value & $N_{\text{total}}$ &
$\sigma_{\uparrow}/\sigma_{\downarrow}$ \\ \hline  \endfirsthead
\hline \endfoot  \hline \hline \endlastfoot
1-74 & -1.9710(44) & $59\times 10^{10}$ & 1.32 \\
75-78, 82-83, 93-94, 101, 133 & -2.0858(26) & $19\times 10^{10}$ & 1.39 \\
79, 89, 104, 116 & 9.2853(15) & $64\times 10^{9}$ & 1.34 \\
80-81, 84, 92, 105-106, 117-118, 131-132 & -7.3999(19) & $12\times 10^{10}$ & 1.35 \\
85-86 & 0.91509(81) & $29\times 10^{9}$ & 1.40 \\
88, 113 & -0.03943(45) & $11\times 10^{9}$ & 1.24 \\
91, 114 & -1.21525(47) & $10^{10}$ & 1.28 \\
95-96, 107-108, 120-121, 125, 134-139, 141-142, 144-148 & 11.6975(35) & $30\times 10^{10}$ & 1.14 \\
97-98 & 0.07633(84) & $30\times 10^{9}$ & 1.77 \\
103, 115 & -0.21851(49) & $13\times 10^{9}$ & 1.50 \\
110, 124 & -1.67843(52) & $13\times 10^{9}$ & 1.35 \\
119, 122, 140, 143 & -10.6235(17) & $69\times 10^{9}$ & 1.20 \\
127-128 & -2.10043(82) & $29\times 10^{9}$ & 1.34 \\
129-130 & 1.17276(61) & $17\times 10^{9}$ & 1.34 \\
149-168 & -0.6220(46) & $44\times 10^{10}$ & 1.08 \\
169-170 & -0.20117(59) & $15\times 10^{9}$ & 1.38 \\
172, 175 & 0.03046(39) & $87\times 10^{8}$ & 1.22 \\
173, 180 & -0.5121(10) & $38\times 10^{9}$ & 1.53 \\
176, 179 & 0.24323(93) & $33\times 10^{9}$ & 1.34 \\
177-178 & 0.94064(71) & $20\times 10^{9}$ & 1.29 \\
183, 208, 212, 219 & 18.2163(17) & $89\times 10^{9}$ & 1.28 \\
185, 195 & -0.52238(43) & $10^{10}$ & 1.36 \\
186, 199, 209, 213 & -6.8978(17) & $91\times 10^{9}$ & 1.25 \\
188, 198 & -1.35354(78) & $28\times 10^{9}$ & 1.30 \\
190, 201 & -0.11069(69) & $21\times 10^{9}$ & 1.31 \\
193, 215 & -3.73267(70) & $25\times 10^{9}$ & 1.48 \\
196, 210-211, 216 & -7.9473(14) & $59\times 10^{9}$ & 1.26 \\
202, 214, 217, 220-222 & -0.8907(22) & $13\times 10^{10}$ & 1.05 \\
204, 207 & 1.43937(67) & $21\times 10^{9}$ & 1.35 \\
205, 218 & 0.00898(64) & $19\times 10^{9}$ & 1.36 \\
223-224, 241 & -0.4627(14) & $53\times 10^{9}$ & 1.36 \\
225, 233 & -0.09888(69) & $18\times 10^{9}$ & 1.56 \\
226, 229, 242-243 & 0.5793(14) & $57\times 10^{9}$ & 1.38 \\
227, 230, 247, 250-252 & 0.9197(24) & $12\times 10^{10}$ & 1.08 \\
228, 238 & -0.42075(69) & $21\times 10^{9}$ & 1.40 \\
231-232, 248-249 & -0.3857(13) & $44\times 10^{9}$ & 1.28 \\
235, 237 & -1.40923(60) & $15\times 10^{9}$ & 1.52 \\
239-240 & 1.01660(76) & $25\times 10^{9}$ & 1.25 \\
244-246 & 0.30280(80) & $24\times 10^{9}$ & 1.10 \\
260, 265 & 1.25169(40) & $10^{10}$ & 1.32 \\
262, 266 & 5.27326(83) & $28\times 10^{9}$ & 1.42 \\
264, 267-268 & -10.52672(91) & $34\times 10^{9}$ & 1.21 \\
\end{longtable}
\normalsize

\small
\begin{longtable}{ccccc}\caption{Contributions of the gauge invariant classes $(k,m,m')$ to $A^{(8)}_1$, a comparison of these results with the semianalytical values from ~\cite{laporta_8} (continued)} \\
\hline \hline Class & Value & Semianalytical value &
$N_{\text{total}}$ & $\sigma_{\uparrow}/\sigma_{\downarrow}$ \\
\hline  \endhead
\caption{Contributions of the gauge invariant classes $(k,m,m')$ to $A^{(8)}_1$, a comparison of these results with the semianalytical values from ~\cite{laporta_8}}\label{table_gauge_4loops} \\
\hline \hline Class & Value & Semianalytical value &
$N_{\text{total}}$ & $\sigma_{\uparrow}/\sigma_{\downarrow}$ \\
\hline  \endfirsthead \hline \endfoot  \hline \hline \endlastfoot
$(1,3,0)$ & -1.9710(44) & -1.9710756168358 & $59\times 10^{10}$ & 1.32 \\
$(2,2,0)$ & -0.1415(56) & -0.1424873797999 & $96\times 10^{10}$ & 1.26 \\
$(1,2,1)$ & -0.6220(46) & -0.6219210635351 & $44\times 10^{10}$ & 1.08 \\
$(3,1,0)$ & -1.0424(44) & -1.0405424100126 & $70\times 10^{10}$ & 1.23 \\
$(2,1,1)$ & 1.0842(37) & 1.0866983944758 & $38\times 10^{10}$ & 1.21 \\
$(4,0,0)$ & 0.5120(17) & 0.512462047968 & $13\times 10^{10}$ & 1.28 \\
\end{longtable}
\normalsize

\subsection{Technical information}\label{subsec_tech}

Table \ref{table_tech} contains a summary of results and a technical
information. The meanings of the fields $N_{\text{total}}$,
$N_{\text{EIA}}^{\text{fail}}$, $N_{\text{IA}}^{\text{fail}}$,
$N_{\text{128}}^{\text{fail}}$,
$\triangle_{\text{EIA}}^{\text{fail}}$,
$\triangle_{\text{IA}}^{\text{fail}}$,
$\triangle_{\text{128}}^{\text{fail}}$ are defined in Section
\ref{subsec_indi}. The GPU performance\footnote{The announced by
NVidia peak performance of one GPU of NVidia Tesla K80 for the
double precision is $1.45$ TFlops.} for these computations is
measured in floating point operations per second (Flop/s) and
interval operations per second (Interval/s) in the sense of Section
\ref{subsec_ia}, G=Giga, M=Mega.

\tiny
\begin{longtable}{cccccc}\caption{Summary of the results, comparsion with the known (semi)analytical results, technical information (continued)} \\
\hline \hline  & 2 loops & 3 loops & 4 loops & 5-loop ladder &
6-loop ladder \\ \hline  \endhead
\caption{Summary of the results, comparsion with the known (semi)analytical results, technical information}\label{table_tech} \\
\hline \hline  & 2 loops & 3 loops & 4 loops & 5-loop ladder &
6-loop ladder \\ \hline  \endfirsthead \hline \endfoot  \hline
\hline \endlastfoot
Value & $-0.3441651(34)$ & $0.90485(10)$ & $-2.181(10)$ & $11.6530(58)$ & $34.31(20)$ \\
(Semi)analytical value for comparison & $-0.344166387$ & $0.904979$ & $-2.1769$ & $11.6592$ & $34.367$ \\
References for the (semi)analytical value & \cite{analyt2_p} & \cite{analyt_b,analyt_e,analyt_d,analyt_c,analyt_f,analyt3} & \cite{laporta_8} & \cite{caffo} & \cite{caffo} \\
$\sigma_{\uparrow}/\sigma{\downarrow}$ & $1.02$ & $1.05$ & $1.21$ & $1.16$ & $1.74$ \\
$N_{\text{total}}$ & $33\times 10^{11}$ & $81\times 10^{11}$ & $32\times 10^{11}$ & $29\times 10^{9}$ & $10^{10}$ \\
$N_{\text{EIA}}^{\text{fail}}$ & $71\times 10^{8}$ & $17\times 10^{10}$ & $18\times 10^{10}$ & $32\times 10^{8}$ & $12\times 10^{8}$ \\
$N_{\text{IA}}^{\text{fail}}$ & $68\times 10^{6}$ & $21\times 10^{8}$ & $13\times 10^{8}$ & $90\times 10^{5}$ & $72\times 10^{5}$ \\
$N_{\text{128}}^{\text{fail}}$ & $2$ & $12590$ & $77775$ & $934$ & $4504$ \\
$\triangle_{\text{EIA}}^{\text{fail}}$ & $0.002$ & $0.4$ & $2$ & $5$ & $20$ \\
$\triangle_{\text{IA}}^{\text{fail}}$ & $0.0001$ & $0.002$ & $0.2$ & $0.4$ & $3$ \\
$\triangle_{\text{128}}^{\text{fail}}$ & $-2\times 10^{-19}$ & $-10^{-6}$ & $-0.0006$ & $4\times 10^{-10}$ & $-5\times 10^{-5}$ \\
Total calculation time & 21 h 37 min & 5 d 8 h & 7 d & 4 h 38 min & 8 h 24 min \\
Share in the time: double-precision EIA & $19.1\%$ & $41.7\%$ & $54.5\%$ & $56.4\%$ & $42.0\%$ \\
Share in the time: double-precision IA & $0.1\%$ & $1.6\%$ & $9.1\%$ & $15.4\%$ & $24.4\%$ \\
Share in the time: 128-bit mantissa IA & $0.2\%$ & $2.7\%$ & $9.2\%$ & $6.7\%$ & $24.3\%$ \\
Share in the time: 256-bit mantissa IA & $0.0\%$ & $0.3\%$ & $2.1\%$ & $8.1\%$ & $5.2\%$ \\
Share in the time: sample generation & $63.7\%$ & $45.9\%$ & $21.7\%$ & $12.0\%$ & $3.7\%$ \\
Share in the time: other operations & $16.9\%$ & $7.7\%$ & $3.4\%$ & $1.3\%$ & $0.3\%$ \\
GPU speed: double-precision EIA, GFlop/s & $334.24$ & $222.72$ & $234.26$ & $187.93$ & $292.67$ \\
GPU speed: double-precision EIA, GInterval/s & $53.76$ & $63.51$ & $142.27$ & $103.04$ & $240.91$ \\
GPU speed: double-precision IA, GFlop/s & $254.11$ & $221.41$ & $255.85$ & $249.00$ & $287.94$ \\
GPU speed: double-precision IA, GInterval/s & $36.23$ & $35.80$ & $47.22$ & $45.60$ & $55.81$ \\
GPU speed: 128-bit mantissa IA, GFlop/s & $0.81$ & $1.59$ & $1.58$ & $1.63$ & $1.66$ \\
GPU speed: 128-bit mantissa IA, GInterval/s & $0.11$ & $0.23$ & $0.26$ & $0.30$ & $0.32$ \\
GPU speed: 256-bit mantissa IA, MFlop/s & $0.0204$ & $0.0881$ & $0.3503$ & $0.1378$ & $4.8504$ \\
GPU speed: 256-bit mantissa IA, MInterval/s & $0.0028$ & $0.0124$ & $0.0537$ & $0.0252$ & $0.9401$ \\
Integrand code size: not compiled & 887 KB & 31 MB & 2.5 GB & 23 MB & 186 MB \\
Integrand code size: compiled & 12 MB & 115 MB & 4 GB & 34 MB & 252 MB \\
\end{longtable}
\normalsize

\section{CONCLUSION}\label{sec_conclusion}

The method for numerical evaluation of $A_1^{(2n)}[\text{no lepton
loops}]$ described in ~\cite{volkov_2015,volkov_prd} was
significantly improved. The main improvements are:
\begin{itemize}
\item probability density functions for Monte Carlo integration giving
a better convergence;
\item a method of Monte Carlo error estimation;
\item a method of high-speed arithmetic calculations with round-off
error control;
\item a realization on high-speed graphics processors.
\end{itemize}
The values for $n=2,3,4$ were obtained and compared with the known
analytical and semianalytical ones as well as the contributions of
the 5-loop and 6-loop ladder graphs. The results were presented in a
form allowing to check them by parts using another methods. The
2-loop and 3-loop contributions were compared with the known values
in detail, the 4-loop ones were compared for 6 gauge invariant
classes. All obtained results are in good agreement with the known
ones. The results showed that the developed method and its
realization allows us to obtain high-precision values for high-order
QED contributions to $a_e$ even without appealing to supercomputers.

The ability of using nonadaptive Monte Carlo algorithms for
obtaining high-precision results was verified. The behavior of the
Monte Carlo samples was analyzed in detail. The necessity of
probability distribution extrapolation for obtaining correct error
estimations was explained, the method was presented. The impact of
possible round-off errors was investigated in detail, the necessity
of controlling them and applying high-precision arithmetic was
justified. The developed high-speed method of controlling round-off
errors can be used for other calculations in Quantum Field Theory
that are based on numerical subtraction of divergences under the
integral sign.

The performed 6-loop calculation showed a big impact of
high-precision arithmetic to the calculation speed and the necessity
of accurate error estimation, but the 3-loop calculation discovered
a sensitivity to a selection of a pseudorandom generator.

The realization on GPU showed a very good performance. For example,
the speed of obtaining integrand values was improved by 3000 times
in comparison with ~\cite{volkov_prd} for the 5-loop ladder graph.

In closing, let us recapitulate some theoretical problems that still
remain open:
\begin{enumerate}
\item to prove mathematically (or disprove) that the developed
subtraction procedure leads to finite integrals for any Feynman
graph for any order of the perturbation series;
\item to create a mathematical foundation for the probability density
functions that were used for the Monte Carlo integration;
\item to generalize the concept of I-closure and to develop a method
of obtaining $\ffdeg(s)$ for graphs with lepton loops;
\item to explain why the contributions of gauge invariant classes
are relatively small, but the contributions of individual graphs or
even sets from Section \ref{subsec_classes} are relatively large; is
it true for the higher orders of the perturbation series?
\end{enumerate}

\section{ACKNOWLEDGEMENTS}\label{sec_ack}

The author thanks Andrey Kataev for interesting discussions and
helpful recommendations, Andrey Arbuzov for his help in
organizational issues, Predrag Cvitanovi\'{c} for fruitful
discussion and inspiring ideas, Ivan Krasin for his help in
understanding NVidia graphics accelerators and Google services, and
Denis Shelomovskij for his help in D programming issues. Also, the
author is very pleased that Google provides an ability to rent
computers with powerful graphics accelerators for free without any
bureaucracy. Special thanks are due to the reviewers for careful
reading of the article and valuable advices.

\end{document}